\documentclass[12pt]{article}
\usepackage{amssymb,amsmath} 
\usepackage[pdftex]{graphicx}
\usepackage{bbm}
\usepackage{bm}% bold math
\newcommand{\Letter}{
  \setlength{\textwidth}{7 in}
   \setlength{\textheight}{9.0in}
  \hoffset=-.75in
   \voffset=-1.15in }
\def\baray{\begin{eqnarray*}}
\def\earay{\end{eqnarray*}}
\def\ba{\begin{eqnarray}}
\def\ea{\end{eqnarray}}
\def\be{\begin{equation}}
\def\ee{\end{equation}}
\def\ben{\begin{equation} \nonumber}
\def\een{\end{equation}}

\Letter

\begin{document}

\begin{titlepage}
\begin{center}

{\hbox to\hsize{
\hfill
arXiv:0903.4686
}}

{\hbox to\hsize{
\hfill
NSF-KITP-09-36
}}

{\hbox to\hsize{
\hfill
MIFP-09-16
}}

\vspace{2cm}

{\large \bf Gravitational Waves from Broken Cosmic Strings:  \\
The Bursts and the Beads\\[2mm]}

\vspace*{1.0cm}

{Louis Leblond\footnote{Electronic address: lleblond@physics.tamu.edu}, Benjamin Shlaer\footnote{Electronic address: shlaer@cosmos.phy.tufts.edu}, and Xavier Siemens\footnote{Electronic address: siemens@gravity.phys.uwm.edu}\\[8mm]

{${}^1$ \it Mitchell Institute for Fundamental Physics, Department of Physics, Texas A\&M University, College Station, Texas, 77840, USA\\
and\\
Kavli Institute for Theoretical Physics, University of California, Santa Barbara, CA,
93106, USA}\\
[2mm]
{${}^2$ \it Institute of Cosmology, Department of Physics and Astronomy\\Tufts University, Medford, MA  02155, USA}\\[2mm]
{${}^3$ \it Center for Gravitation and Cosmology, Department of Physics,
P.O.  Box 413, University of Wisconsin - Milwaukee, Wisconsin, 53201, USA}}

\vspace*{1.5cm}
\date{\today}

%%%%%%
{\bf Abstract}\\
\end{center}

We analyze the gravitational wave signatures of a network of
metastable cosmic strings.  We consider the case of cosmic string
instability to breakage, with no primordial population of monopoles.
This scenario is well motivated from GUT and string theoretic models
with an inflationary phase below the GUT/string scale.  The network
initially evolves according to a scaling solution, but with breakage
events resulting from confined monopoles (beads) being pair produced and
accelerated apart.  We find these ultra-relativistic beads to be a potent source of gravitational waves bursts,
detectable by Initial LIGO, Advanced LIGO, and LISA. Indeed, Advanced LIGO could observe bursts
from strings with tensions as low as $G\mu \sim 10^{-12}$. In addition, we find that
ultra-relativistic beads produce a scale-invariant stochastic background detectable by LIGO,
LISA, and pulsar timing experiments. The stochastic background is scale invariant up to
Planckian frequencies. This phenomenology provides new constraints and signatures of cosmic
strings that disappear long before the present day.

%\end{center}
\noindent

%\vfill
%\hrulefill\hspace*{4in}

%{\footnotesize $^{\dagger}$ Email:
%\parbox[t]{7in}{}}

%\tableofcontents
%\newpage
\end{titlepage}
\newpage
\tableofcontents

\section{Introduction}

Cosmic strings can be produced in any phase transition where the
vacuum group manifold is not simply connected, {\em e.g.}, during the
breaking of a $U(1)$ symmetry.  As topological defects, they appear to
be completely stable at the level of the effective action where the
$U(1)$ symmetry is manifest.  However, stability may break down in the
ultraviolet.  If, for example, the $U(1)$ symmetry is embedded into a
non-abelian gauge group at some higher energy scale, there must exist
monopoles somewhere in the spectrum.  The flux from these monopoles is
confined under the $U(1)$ symmetry breaking, and so hybrid defects can
exist with strings ending on beads (we distinguish between monopoles and beads: the latter
have their flux confined into strings). A second source of instability
for global defects can arise from non-perturbative effects such as
instantons, which generically lift the vacuum degeneracy to isolated
points.  The cosmic strings then become boundaries of domain walls.
Both instabilities can lead to a rapid demise of the network,
rendering cosmic strings metastable at best.  The question of cosmic
string stability in the context of hybrid defects was originally
investigated in Grand Unified Theories (GUT).  The Langacker-Pi
mechanism \cite{Langacker}, whereby monopoles created at the GUT phase
transition have their flux confined into strings, was proposed to
solve the monopole problem.  It eliminated the monopoles by confining
their flux into cosmic strings, which would in turn oscillate and quickly decay to
radiation. In this paper, we will be interested in looking at the phenomenological signals 
of such a decaying network of cosmic strings.

Cosmic strings formed in the early universe could result in a myriad of astrophysical phenomena.
These include ultra-high energy cosmic rays~\cite{bands}, gamma ray bursts~\cite{berez}, radio
wave bursts~\cite{Vachaspati:2008su}, magnetogenesis~\cite{Battefeld:2007qn}, strong
lensing~\cite{Mack:2007ae,Gasparini:2007jj,Christiansen:2008vi}, weak
lensing~\cite{Dyda:2007su}, microlensing~\cite{Chernoff:2007pd,Kuijken:2007ma}, along with
effects on the cosmic microwave background (CMB) polarization~\cite{Baumann:2008aj}, the CMB
spectrum at small angular scales~\cite{Fraisse:2007nu,Pogosian:2008am}, and the cosmic 21~cm
power spectrum~\cite{Khatri:2008zw}.

One of the most exciting recent predictions is the observability of cosmic strings via their
gravitational wave signals. Cosmic string produce powerful bursts of gravitational radiation
that could be detected by interferometric gravitational wave detectors such as LIGO, Virgo, and
LISA~\cite{DV1,SCMMCR}. In addition, the stochastic gravitational wave background can
be detected or constrained by various observations including Big Bang Nucleosynthesis (BBN),
pulsar timing experiments~\cite{Jenet:2006sv}, and interferometric gravitational wave
detectors~\cite{DV1,SMC}.

An appealing source of cosmic strings from string-theoretic models such
as brane inflation has led to a renewal of activity in recent years
\cite{Tye}.  In such a scenario, the inflaton is represented by the
relative location of a D-brane within the compactified
extra dimensions.  In some models, inflation ends with brane-antibrane
annihilation, and the resulting open-string tachyon condensation leads
to the formation of co-dimension two topological defects, as is
typical in hybrid-inflationary models.  In the generic case, a network
of cosmic -F and/or -D strings is produced.  The phenomenology of a
network of stringy cosmic strings is richer than the standard abelian
Higgs cosmic strings due to new parameters such as the spectrum of
tensions \cite{CMP, Firouz}, intercommutation probability \cite{Jack}
(see \cite{Jackson:2009fk} for discussion of bursts from intercommutation)
lensing \cite{Shlaer:2005ry}, and as we shall discuss here, a decay rate
per unit length.

While the production of cosmic strings is essentially guaranteed
following brane anti-brane annihilation (or more general hybrid
inflationary models), the stability of such strings is model
dependent.  In the context of heterotic string theory, it was realized
early on that the strings would be unstable due to the formation of
domain walls stretching between various parts of the network
\cite{Witten} (the tension was also too high in these early models).
The situation was reanalyzed in the context of Type IIB flux
compactification first by Copeland, Myers and Polchinski \cite{CMP}
and another source of instability, to breakage, was discussed.  These
two instabilities are complimentary, in the sense that typically exactly
one will occur.  This was further analyzed in \cite{LebTye} where it
was shown that the generalized Green-Schwarz mechanism eliminates the domain wall
instability leaving only the possibility of breakage on beads.
The reason breakage is merely a possibility is because it is often not
clear what the monopole is (or even if it exists) at the level of the
low energy effective action.  
Multiple examples of
stringy monopoles are known, ranging from BI-ons \cite{Callan} to
configurations of branes (for a recent example see \cite{Verlinde}).
In some cases, the strings carry a fraction of the minimum magnetic charge,
and the network of strings, instead of having endpoints, has multiple
strings ending on a given bead.  (An
important example is the ``baryon" D3 branes described in
\cite{Gubser}).  This leads to so-called cosmic necklaces, and they 
have interesting phenomenology of their own \cite{LebWym, Blanco, BlancoPillado:1999cy, Berezinsky:1997td}.

In this paper, we will be interested in the scenario where the strings
are unstable to breakage due to monopoles/beads.  We will not rely on any
specific string-theoretic models, and we will assume vanilla cosmic
strings and beads;  they are
neutral under every long range force except gravity.

Group theoretic considerations require that the symmetry breaking
scheme must produce monopoles at a higher scale than the subsequent
cosmic strings.  The pattern of symmetry breaking is
\cite{PreskillVilenkin} (see also \cite{Jeannerot:2003qv} for more discussion on cosmic strings in GUT models)
\be
G \rightarrow H\times U(1) \rightarrow H \; ,
\ee
where $G$ is a semi-simple Lie group.  Monopoles are created under the
first symmetry breaking, and their flux is confined under the second, forming a
network of beads and strings. 
A corollary of such a sequence is that the cosmic strings are at best
metastable.  Such a scenario has previously been considered \cite{Martin:1996ea, Vilenkin:1982hm},
and it was found that the beads rapidly annihilate due to the confining
potential generated by the strings.  The close proximity of beads
means that such a network is {\em unstable}.  However, in the special
case where the monopole creating phase transition occurs before
inflation and the string production phase after inflation, there will not
be any primordial population of beads.\footnote{In brane inflation models,
cosmic strings are naturally produced at the end of inflation, while
monopoles are not \cite{Tye}.}  Such a network is merely {\em
  metastable}.  Strings will then only decay via the Schwinger
process (see \cite{Monin:2008mp}), whose time scale is given by the decay rate per unit length
\be\label{eq:rate}
\Gamma_2 = \frac{\mu}{2\pi}\exp\left(-\pi m^2/\mu\right)\; ,
\ee
where $\mu$ is the string tension and $m$ is the bead mass.  We
parametrize the bead mass in units of the string tension by the
parameter $\kappa = m^2/\mu$.  From the symmetry breaking pattern
above $\kappa \gtrsim 1$.  A derivation of Eq.~(\ref{eq:rate}) is
given in Appendix \ref{sec:rate}.  For large values of $\kappa$ (for
example, $\kappa > 86$ if the tension is $G\mu\sim 10^{-7}$), the 
cosmic string lifetime is greater than the age of the universe, and
the network is stable for all practical purposes.  For small values of
$\kappa$, the network decays almost immediately upon formation.  
%In this paper, we investigate the gravitational waves signatures, both burst
%and stochastic, left by any such decaying network of cosmic strings.

Martin and Vilenkin first studied the gravitational waves produced by
these hybrid defects \cite{MV}.  They calculated the power radiated
per frequency interval after solving the Nambu-Goto equations of the
string/bead system.  They assumed a network with a nonzero primordial
monopole $\to$ bead population and calculated the total power and stochastic
background of the resulting gravitational waves.  In this paper we
extend their calculation to include an analytic solution of the
waveform in different frequency intervals, and we consider the string motivated
scenario in which the initial population of monopoles is absent due to
inflation.

We also calculate the phenomenology of highly focused gravitational
wave bursts.  In a celebrated paper \cite{DV1}, Damour and Vilenkin
have shown that a network of cosmic strings can be observed via
experiments like LIGO and LISA through the gravitational wave bursts
produced by cusps and kinks on string loops.  We find that intense
bursts are produced by the acceleration of the ultra-relativistic
monopoles, and we present the range of parameters where we can hope to
detect these bursts.  This calculation is of theoretical interest in
its own right since a cosmic string ending on beads is the
archetype of gravitational radiation emitted from a linearly
accelerated mass, a problem famous for its simplicity and paradoxes,
as we will review.

Our main results are that for a moderate tuning of the monopole mass, 
Advanced LIGO will be able to detect strings whose
tension exceeds $G\mu \sim 10^{-12}$.  
We also find that a stochastic
signal persists for networks which decay at any time between reheating
and the present day.  Because loop production is highly suppressed when
the typical length of segment becomes sub-horizon, the gravitational wave 
phenomenology from an unstable network is complimentary to that of stable 
cosmic strings: an initial ``stable string" epoch of loops is usurped by oscillating segments, followed
eventually by the complete demise of the string network.

The spectrum is very nearly independent of both frequency
and $\kappa = m^2/\mu$.  The frequency cutoffs range from the Hubble scale to well beyond the Planck scale.  This is an indication
that massive ({\em e.g.,} dilaton) radiation may contribute to the phenomenology \cite{Damour:1996pv,Babichev:2005qd}.

In \S\ref{metastable}, we review the issue of metastability,
with particular emphasis on the recent models motivated by string
theory.  In \S\ref{waveform}, we calculate the gravitational
waveform and power radiated by strings ending on massive beads.  In
\S\ref{cosmology}, we track the cosmology of such a network and
find the density and length distribution of cosmic strings.  We then
calculate the observational signals from such a network in 
\S\ref{results} before concluding.  We have relegated many of the more
technical details to a series of appendices.

\section{Metastable Cosmic Superstrings}
\label{metastable}
\begin{figure}[h] %  figure placement: here, top, bottom, or page
   \centering
   \includegraphics[width=7in]{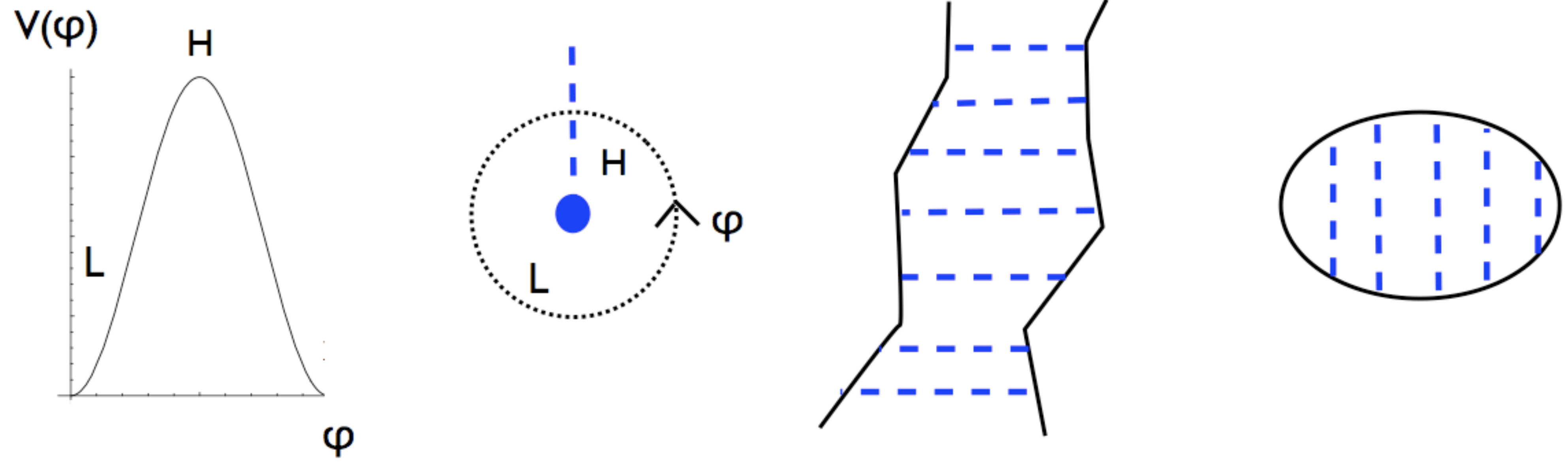}
   \caption{As the axion winds around the string, the non-perturbative potential $V(\varphi)$ must attain its maximum in some direction.  A domain wall thus confines any string.  Such walls will pull infinite strings towards each other and cause rapid demise of string loops.} 
   \label{domwal}
\end{figure}

Cosmic strings can decay by either becoming the boundary of a domain wall (the strings are {\em confined}) or by developing boundaries themselves through nucleation of bead endpoints (the strings are {\em screened}).  Because the boundary of a boundary does not exist, these two types of instabilities are mutually exclusive. To see how they arise, let us start with a global cosmic string whose dynamics is well described outside the core by the Kalb-Ramond action
\be\label{eq:kalb}
S = - \mu \int_{\Sigma} d^2\sigma + \eta \int_{\Sigma} B_2 - \int_{\mathbb{R}^{3,1}}\hspace{-2mm} d^4x\,  \tfrac{1}{12}H_{\mu\nu\rho}H^{\mu\nu\rho}\; ,
\ee
where $\Sigma$ is the string worldsheet, $B_2$ is a 2-form with $H_3=dB_2$ the associated field strength, $\mu$ is the string tension and $\eta$ its charge.  We have suppressed the Einstein-Hilbert term.  In 3+1 dimensions, the massless 2-form is dual to an axion: $dB_2 =  \star d\phi$. The Kalb-Ramond action (Eq.~\ref{eq:kalb}) therefore represents a cosmic string with a magnetically sourced long-range axionic force.  There is a global $U(1)$ shift symmetry of the axion $\phi \rightarrow \phi +c$, and since this string carries a global charge, local breakage is forbidden by topological considerations.  Interestingly this fails to protect the strings, as domain walls will confine and rapidly eliminate them \cite{VilenkinShellard,CMP}.  This occurs when instantons generate a periodic potential for the axion of the form $V(\phi) = \Lambda^4(1+\cos\phi)$ where $\Lambda$ is some non-perturbative energy scale.  At the minimum of the instanton potential, the global $U(1)$ is broken to a discrete subgroup, and domain walls arise.  As the axion winds around the global strings, it will necessarily have ``to go over the bump" of the periodic potential; therefore the string is always going to be the boundary of a domain wall (see Fig.~\ref{domwal}).  In general, the wall's tension can be high and will confine the strings together. Hence we expect a rapid demise of the string-domain wall network.  The phenomenology of such an hybrid network may be interesting, and so we leave this for future work.  

We will be interested in the case of a local string which has no long range axionic force.  The abelian Higgs model is the archetype of such a local cosmic string, but it is illuminating to minimally generalize the Kalb-Ramond action to include a gauge field.  Consider the action
\be\label{stuck}
S = - \mu \int_{\Sigma} d^2\sigma + \eta \int_{\Sigma} B_2 - \int_{\mathbb{R}^{3,1}}d^4x \left( \tfrac{1}{12}H_{\mu\nu\rho} H^{\mu\nu\rho} + \tfrac{1}{4}F_{\mu\nu}F^{\mu\nu}\right) + \xi B_2\wedge F_2\; ,
\ee 
 where $F_2=dA$ is the $U(1)$ field strength and the last term is the coupling between the axion and the gauge field parametrized by $\xi$.  This can be shown to be equivalent to the abelian Higgs model in the limit where the massive Higgs is integrated out of the theory.  The $B_2\wedge F_2$ spontaneously breaks the $U(1)$ gauge symmetry and the gauge field picks up a mass $\xi$.  This is just the St\"uckelberg model. (In string theory, these types of terms can implement the generalized Green-Schwarz mechanism for anomaly cancellation).  The massive gauge field has eaten the axion, and it now propagates three degrees of freedom.  There is no long range axionic interaction, and the charge is therefore screened.  Gauge invariance forbids an instanton potential for the axion, and it again appears as if the string is completely stable. 

This is no longer generically true when we consider the possible UV completion of the theory.  Indeed, the $U(1)$ could have been embedded in a non-abelian gauge group at some higher scales, in which case heavy monopoles will exist for the string to break on.  In string theory, more exotic UV completions can also lead to monopoles.  For example, by generalizing the kinetic term of the gauge field (which we took to be canonical 
in Eq.~\ref{stuck}) to the Born-Infield action (a non-linear generalization of Maxwell's theory), one can show the existence of monopole-like solutions representing the endpoint of a string on a brane \cite{Callan}.  That is, even if the gauge group remains $U(1)$ at high energy, a cosmic $(p,q)$-string can still break on a brane in string theory and the endpoints will look like monopoles (or dyons) in four dimensions.  The only way to forbid the existence of monopoles which could serve
as a given string's endpoint is to use the Dirac quantization condition, whereby such a local string is properly classified as a stable Aharonov-Bohm string.

Polchinski has conjectured that monopoles of minimum Dirac charge always exists in string theory \cite{Open}.  This reduces the stability question to that of the existence of particles of fractional electric charge (in units of the abelian Higgs charge).  If no fractional electric charges exist, than the Dirac monopole will carry the same charge as the string (flux tube), and breakage can occur.  If particles exist of electric charge $1/M$, then bead nucleation will only permit $q$-strings to decay into $q - M$ strings, and so only $M$ coincident strings can break completely.  One can thus look at stability in two equivalent ways.  On the one hand, if the necessary magnetic charge for a string to terminate on does not exist, the string cannot break.  To forbid the
existence of such an object in a UV independent way, we simply require that the charge quantization satisfy Dirac's bound
\be
2\pi g_{min} = 1/e_{min}\,\,,
\ee
and then introduce particles of electric charge equal to a fraction of the abelian Higgs charge.
Alternatively, one can show that by introducing particles of fractional electric charge, there will exist
an observable Aharonov-Bohm phase upon transport of these particles around the string at arbitrary distances.  This topological phase forbids breakage (assuming no additional massless fluxes exist) for the same reason that a global string cannot break, namely causality.  Experiments far from the string should not be sensitive to a simultaneous event near the core.  If no experiment can detect the enclosed string, breakage can occur, and so the lifetime of the strings becomes a phenomenological parameter.  In summary, there are three different cases:
\begin{figure}\label{warpedthroat}
\centering{\includegraphics[width=2.4in]{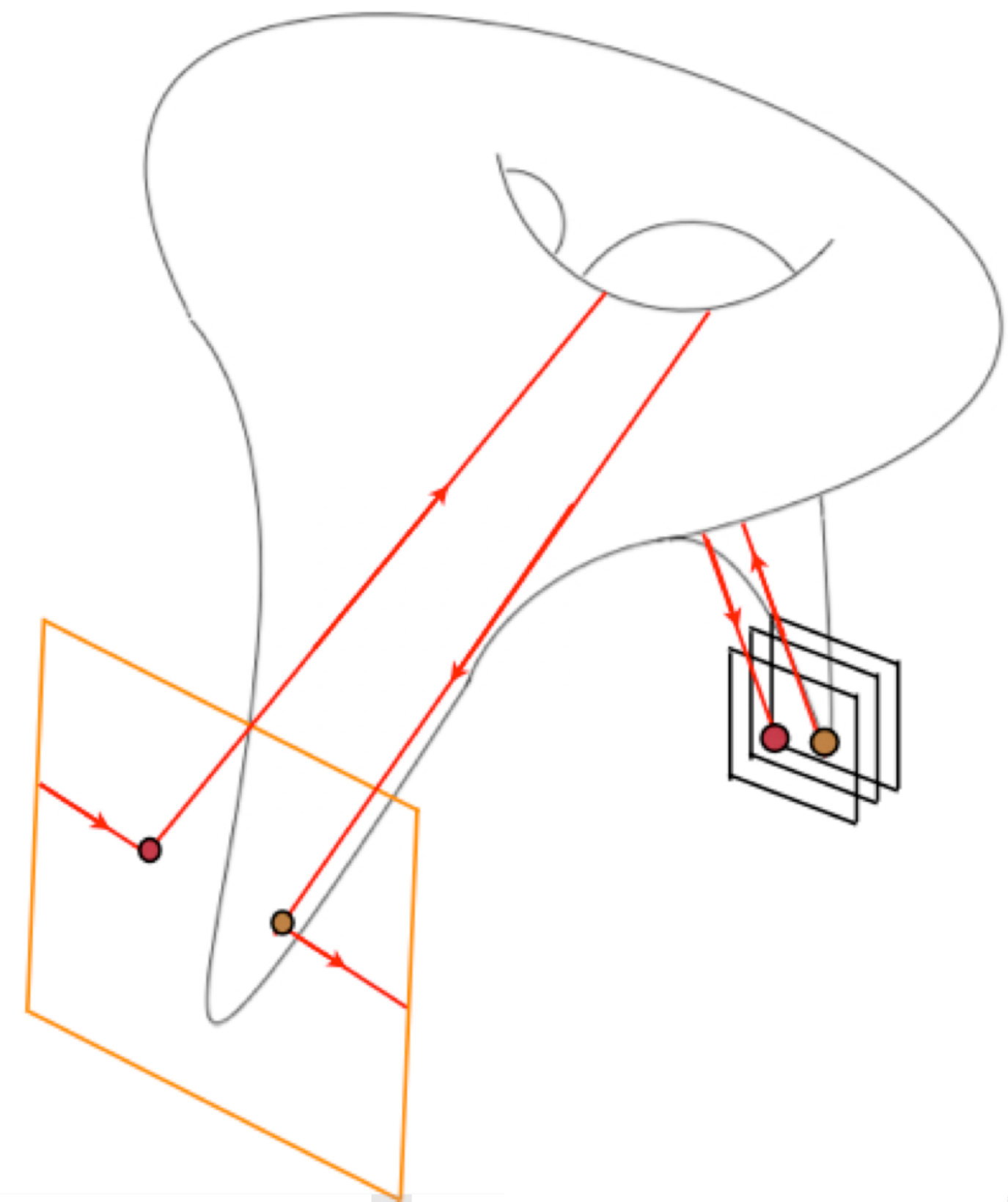}}
\centering{\includegraphics[width=2.4in]{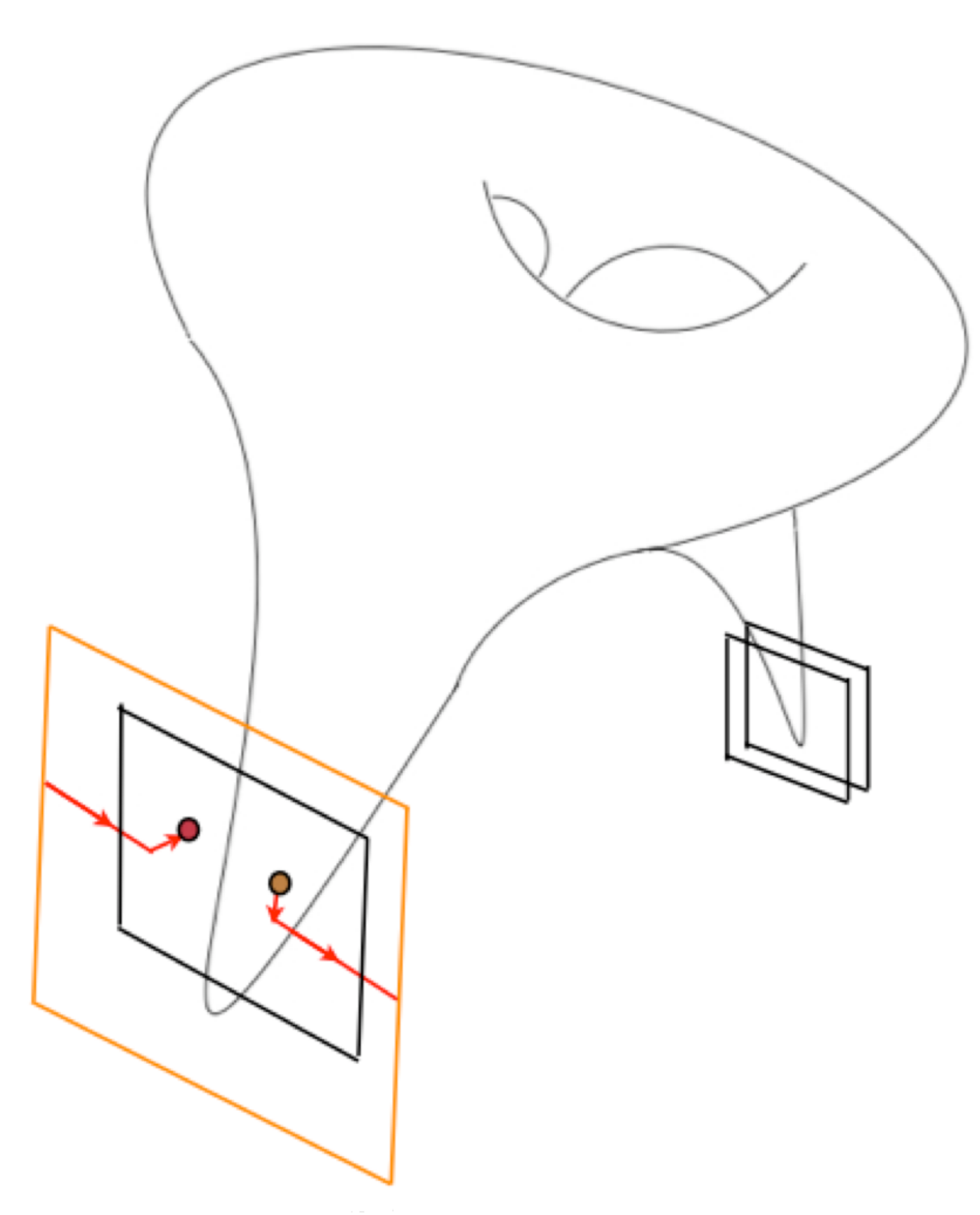}}
\caption{In the context of warped compactifications, different mass scales for the beads/monopoles are possible depending on the specific model. Heavy (bulk scale) beads (left) form when a distant throat is responsible for the instability.  Light beads ($\kappa \gtrsim 1$) form (right) when branes exist near the tip of the inflationary throat.  We assume inflation has already ended in these illustrations.  The orange (dimmer) rectangle represents the far IR, where the strings reside.  Black rectangles are D-Branes.}
\label{warped}
\end{figure}
\begin{itemize}
\item Global strings which are unstable to domain wall formation.
\item Local strings with no Aharonov-Bohm phase, these are not stable due to breakage via monopoles.
\item Local strings with an Aharonov-Bohm phase detectable at infinity. These strings are completely stable.
\end{itemize}
In this paper, we will exclusively discuss the second case where the strings break on monopoles/beads. The lifetime is going to be determined by the mass of these beads. Because this is strongly model dependent, we will simply illustrate a range of possibilities in the context of brane inflation in a warped throat \cite{KKLMMT} (see Fig.~\ref{warped}).  

In this scenario, the cosmic strings are formed at the bottom of a warped throat where the brane and anti-brane annihilate.  If there are spectator branes near the tip after annihilation (which can be stabilized at various positions \cite{D3vacuua}), the string should be able to break directly on them and we expect the mass of the monopoles to be of the same order as the string tension (both set by the warped string scale), and therefore we expect $\kappa \sim \mathcal{O}(1)$ although numerical factors can be quite important\footnote{This was pointed out to us by J.~Blanco-Pillado.}.  If the other branes at the tip are slightly separated, then $\kappa$ can be greater than one, while if the branes are in different throats then we expect $\kappa \gg1$ and a very long-lived network (this case was emphasized in \cite{CMP}).  More exotic monopoles configurations (from various wrapped branes) can also exist in these constructions \cite{Verlinde}.  We expect $\kappa \gtrsim 1$ for these monopoles where the numerical factor depend on the specific setup under consideration; for the rest of this paper we will take it to be a free parameter.  We find that the decay of the network is observable for values of $\kappa \lesssim 80$, which is not necessarily a fine-tuned mass range for the beads.

%%%%%%%%%%%%%%%%%%%%%%%%%%%%%%%%%%%%%%%%%%%%%%
%%%%%%%%%%%%%%%%%%%%%%%%%%%%%%%%%%%%%%%%%%%%%%
%%%%%%%%%%%%%%%%%%%%%%%%%%%%%%%%%%%%%%%%%%%%%%

\section{Gravitational Wave Signal}
\label{waveform}
\begin{figure}[h] %  figure placement: here, top, bottom, or page
   \centering
   \includegraphics[width=3 in]{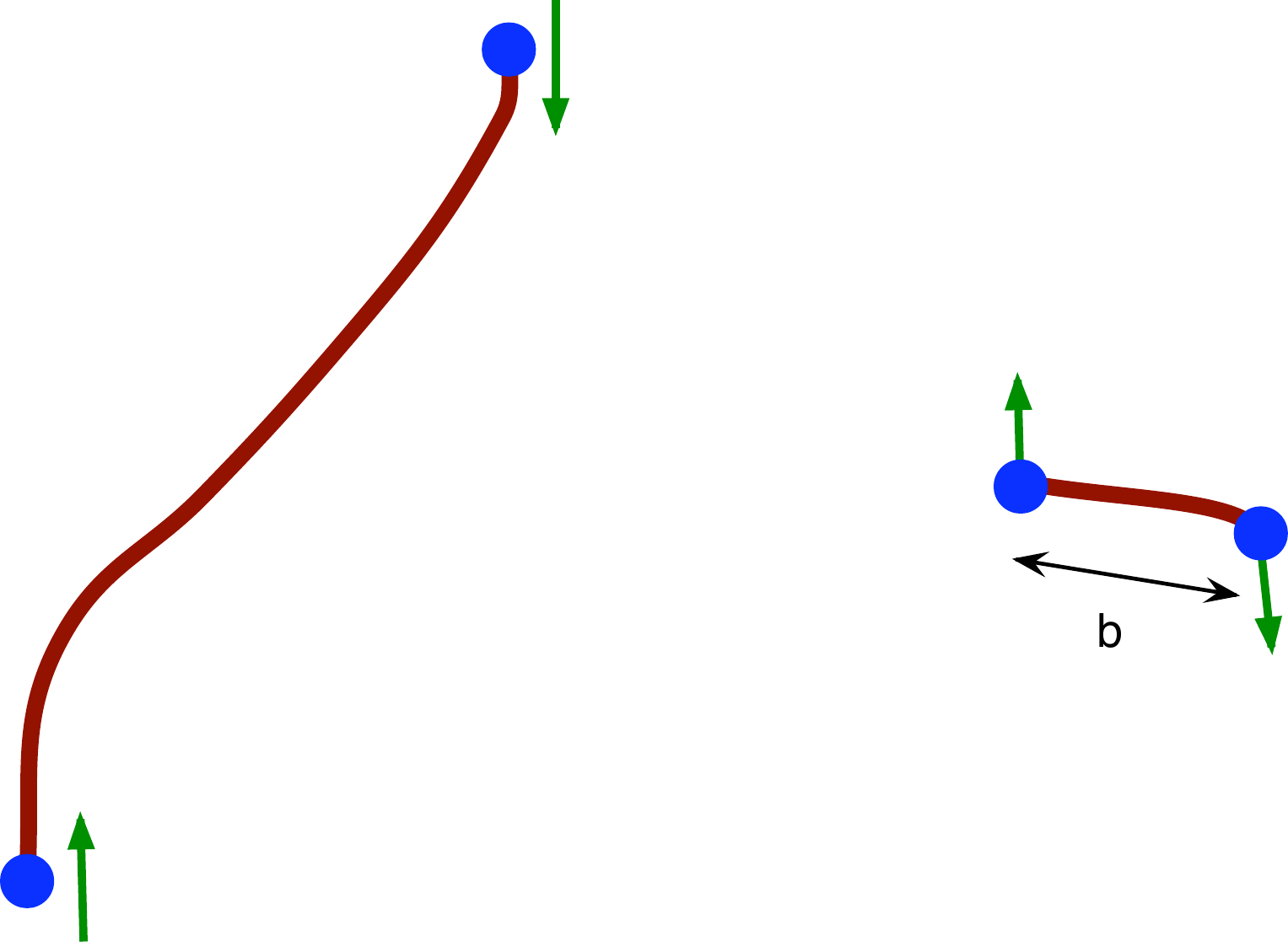} 
   \caption{
An open cosmic string with bead endpoints.  The beads are uniformly
accelerated until the fly-by occurs at some minimum impact parameter
$b$, at which point the acceleration abruptly change sign.  The
presence of kinks (not shown) on the string will also lead to abrupt
change in the acceleration.} 
   \label{monopolestring}
\end{figure}

A straight cosmic string with two bead endpoints will oscillate back and forth once
its length is subhorizon, producing gravitational wave bursts in the process.  This is illustrated in Fig.~(\ref{monopolestring}). 
A cosmic string with endpoints is the archetypical consistent system 
of piecewise uniform acceleration\footnote{In E\&M one can imagine an external force that provides the uniform acceleration. In GR,
this ``external force" will perturb the metric as well and should be included. The solution for a uniformly accelerated mass in GR leads to a class of metric 
called C-metric \cite{Kinnersley:1970zw} (see also \cite{Podolsky:2000pp}). This class of metric always include conical deficit angles, {\em i.e.,} cosmic strings. Therefore, a cosmic string with endpoints is the self consistent way of getting uniform acceleration in general relativity.}.  The beads are being
pulled by the string and accelerate with constant proper acceleration $a=\mu/m$, in the straight string approximation. 
At the fly-by, the acceleration abruptly switches sign, and this discontinuity releases a burst of radiation of characteristic spectrum $1/f^2$.  
Interestingly, this is not responsible for most of the
radiated power.  
It is the ultra-relativistic oscillation 
which produces the most radiated power from the segments, as was first pointed out by 
Martin \& Vilenkin \cite{MV}.  The spectrum from this dominant piece has the scale invariant
$1/f$ form.

\subsection{Radiation from a Uniformly Accelerated Mass}
\label{uniform}
Radiation from uniform acceleration is plagued with several conceptual issues \cite{FUL1, SCH, Boulware, BondiGold, Parrott:1997cp,Candelas}.  
As in electrodynamics, the crux of the matter has to do with a radiation reaction which is proportional to $\dddot{\bf{x}}$ and therefore vanishes for an hyperbolic trajectory.  
This has led some physicists  (such as Pauli) to incorrectly proclaim that there should be no radiation (in the case of electromagnetism) from uniform acceleration,
since it would violate energy conservation.  
The solution to Maxwell's equations for an isolated, uniformly accelerated charge is both simple and singular: the field strength on the light-front is proportional to a delta-function.  It is known
that a particle undergoing uniform acceleration over an arbitrarily long finite proper time {\em does} radiate.  This fact does not undermine conservation of energy or the equivalence principle.  We expect a similar result to hold for gravitational radiation.

In the next section we will compute the spectrum of gravitational radiation from the piecewise
uniformly accelerated beads following 
the work of \cite{MV}. 
This problem has already been studied extensively in the context of E\&M (see for example \cite{NIKI}).
Before going into the detailed calculation, we can summarize all the key features. 
The waveform is a strong $1/f$ burst with the following properties:
\begin{itemize}
\item The power radiated is approximately independent of the length of the segment
\item The power radiated is approximately independent of the mass of the bead
\item The radiation is scale invariant for $1 < f l < \gamma_0^2$
\end{itemize}
The scale invariance of the burst of radiation is a rather unusual feature.  This should be thought
of as a consequence of the ultra-relativistic kinematics.  (A similar behavior was found for the
ultra-relativistic rotating rod by Martin \& Vilenkin.)
Naively, one would expect the discontinuous acceleration to produce
a strong $1/f^2$ burst.  Instead, the even stronger $1/f$ waveform is dominant
whenever $\gamma_0 \gg 1$.

\subsection{Gravitational Waveforms}
For a periodic source with period $T$, the
solution to the linearized Einstein equations in the wave zone has the
form
\begin{equation}
\label{hmunu} 
h_{\mu\nu} ({\bf x},t)=\sum_{n=-\infty} ^{\infty}
\epsilon_{\mu\nu} ^{(n)} ({\bf x},\omega_n) e^{-ik_n \cdot x}\; ,
\end{equation}
where
\begin{equation}
\label{polariz}
 \epsilon_{\mu\nu} ^{(n)} ({\bf x},\omega_n)=\frac{4G}{r} (
T_{\mu \nu}(k_n)-\frac{1}{2} \eta_{\mu \nu} T^\lambda _\lambda(k_n) )\; ,
\end{equation}
is the polarisation tensor, and 
\begin{equation}
\label{TmunuFT1} 
T^{\mu \nu }(k_n)=\frac{1}{T} \int_0 ^{T} dt
\int d^{3}{\bf x} T^{\mu \nu }(x)e^{ik_n \cdot x}\; ,
\end{equation}
is the Fourier transform of the stress-energy tensor. The wave vector is given
by $k^\mu _n=\omega_n(1,{\hat \Omega})$, with $\omega_n=2 \pi n/T$,
and ${\hat \Omega}$ is a unit vector pointing from the source to the
point of observation.  If the bead-string system is confined to the
$z$-axis then the stress-energy tensor only has three non-zero
components: $T^{00}$, $T^{03}$ and $T^{33}$.  The conservation
equations $\nabla_\mu T^{\mu \nu} =0$ in Fourier space is
\begin{equation}
\omega_n T^{0\nu} -k^i T^{i\nu}=0\; ,
\label{conseq} 
\end{equation}
which leaves only one independent component.  

We need the stress-energy tensor of the idealized setup of
a straight string connecting two beads.  This was first calculated by Martin \&
Vilenkin \cite{MV} directly from the Nambu-Goto equations of motion.
 The stress energy tensor
(simplest choice is $T^{03}$) is given in the time-domain by
\begin{equation}
\label{Tmunu01} 
T^{03} (t,{\bf x})=m(\gamma_0 v_0 -a|t|) [\delta ({\bf x}-x_1 (t) {\hat
z}) -\delta ({\bf x}+x_1 (t) {\hat z})]\; ,
\end{equation}
where
\begin{equation}
\label{x1} 
x_i (t) =  (-1)^i \frac{sgn (t)}{a} (\gamma_0 -\sqrt{1+
(\gamma_0 v_0 -a|t|)^2})\; 
\end{equation}
is the position of the two beads (by symmetry $x_1(t) = -x_2(t)$) and $\gamma_0$ and $v_0$ are the maximum gamma factor and speed of
the beads, respectively.  In this case, the period of motion is
$T=\frac{2\gamma_0v_0}{a}$, where $a=\mu/m$ is the proper acceleration
of the beads and the frequency is
\begin{equation}
\omega_n=\frac{n\pi a}{\gamma_0 v_0}\; , 
\label{omega} 
\end{equation} 
The maximum length of the system, when the beads are at rest, is
$l=\frac{2(\gamma_0-1)}{a}$ and so for ultra-relativistic beads we
have, to good approximation, that $T\sim l$.  The Fourier transform of
Eq.~(\ref{Tmunu01}) is,
\begin{equation} 
\label{fft} 
T^{03}(\omega_n ,{\bf k}) = m\gamma_0 v_0 I_n (u)\; ,
\end{equation}
with
\begin{eqnarray}
\label{In} 
I_n (u) &=&  \int_0 ^1 \xi d\xi [\cos (n\pi (1-\xi -\frac{u}{v_0}+u
\sqrt{\xi^2 +1/(\gamma_0 v_0 )^2}))
\nonumber
\\
&& -\cos (n\pi (1-\xi +\frac{u}{v_0}-u\sqrt{\xi^2 +1/(\gamma_0 v_0
)^2}))]
\end{eqnarray}
and $u=k^3/\omega_n$.  We may take the unit vector that defines the
direction of travel of the wave to be
\begin{equation}
\label{Omegavector}
{\hat \Omega}=\left( \begin{array}{c}
\cos\phi \sin\theta\\
\sin\phi \sin\theta\\
\cos\theta\\
\end{array}\right) 
\end{equation}
so that $u=\cos\theta$.  The Fourier transform of the stress-energy
tensor is thus
\begin{equation}
\label{TmunuFT2} 
T^{\mu \nu }(k_n)=
m\gamma_0 v_0 I_n (u)
\left( 
\begin{array}{cccc}
u & 0 & 0 & 1  \\
0 & 0 & 0 & 0 \\
0 & 0 & 0 & 0 \\
1 & 0 & 0 & 1/u\\
\end{array}
\right)
\end{equation}
and its trace is
\begin{equation}
\label{TmunuFT25} 
T^{\lambda} _{\lambda}(k_n)=m\gamma_0 v_0 (u-\frac{1}{u})I_n (u) \; .
\end{equation}
The polarisation tensor reads
\begin{eqnarray}
\label{polariz2}
&&\epsilon_{\mu\nu} ^{(n)} ({\bf x},\omega_n)=
\frac{2Gm\gamma_0 v_0 }{r} I_n(u)
\nonumber
\\
&&
\times\left( 
\begin{array}{cccc}
u+1/u & 0 & 0 & 1  \\
0 & u-1/u & 0 & 0 \\
0 & 0 & u-1/u & 0 \\
1 & 0 & 0 & u+1/u\\
\end{array}
\right).
\nonumber
\\
&&
\end{eqnarray}
As shown in Appendix~\ref{gaugeinv}, the gravitational wave is
linearly polarized and the plus polarization can be written as
\begin{equation}
\label{polariz3cont}
\epsilon_{+}  ({\bf x},\omega)=\frac{2Gm\gamma_0v_0}{r} \frac{u^2-1}{u}I_n(u)\; .
\end{equation}
The waveform in the frequency domain is simply obtained by multiplying
by the period $h(f,r) = T \epsilon_{+}({\bf x},\omega)$.  As was shown
in \cite{MV}, the large frequency ($n\gg \gamma_0^2$) behavior of the
waveform is $1/f^2$.  In Appendix~\ref{continuous}, we re-derive this
result using a continuous approximation that shows clearly that this
tail of the waveform is coming from the discontinuity.  

Almost all of the energy is radiated in the frequency range $1< n < \gamma_0^2$.
A good approximation of the waveform in this regime can be obtained by dropping the second term
in the square root of Eq.~(\ref{In}) and performing the integral
\be
h(f,r) = \frac{16 G m \gamma_0 v_0 l}{r} \frac{\sin^2(n\pi(1-u)/2)}{n^2\pi^2 (u^2-1)}\; .
\ee
Changing variable to $\xi = n\pi(1-u)/2$ gives
\be
h(f,r) =  \frac{4 G m \gamma_0 v_0 l}{r} \frac{\sin^2\xi}{n\pi \xi (1-\xi/(n\pi))}\; .
\ee
The burst is highly focused around $\theta \approx 0$ which corresponds
to small $\xi$.  Together with the fact that $n$ is large, we can drop
the second term in denominator.  All the angular dependence is
therefore describe by $ \frac{\sin^2\xi}{\xi}$ which is zero at $\xi =
[0, \pi, \cdots] $ and has its first maximum at $\xi =\sqrt{2}$.  The
beaming angle can be well approximated by the position of the second
zero at $\xi = \pi$
\be
\theta_f \approx \frac{\mathcal{O}(1)}{\sqrt{n}}\; ,
\ee
This can be written in term of the frequency using $n = f(1+z)l$.  We will consistently refers to $f$
as the observed frequency, while $l$ is the physical (maximal) length
at time of burst.  The waveform at its maximum value is
\be
h(f,r) \approx  C \frac{G\mu}{r} \frac{l}{f}
\label{finvwaveform}
\ee
with $C = \frac{2\sqrt{2} \sin^2\sqrt{2}}{\pi}$.  This is valid for
$\gamma_0 \gg 1$, $v_0 \sim 1$.  Note that there is no redshift
factor $(1+z)$ in this expression.  The waveform swiftly goes to zero for $f<1/T\sim
1/l$ and the approximation we have made breaks down for $n\gg \gamma_0^2$. 
Hence most of the power is radiated in the frequency range
\begin{align}\label{freqrange}
\frac{1}{(1+z)l} < f < \frac{\gamma_0^2}{(1+z)l}\;.
\end{align}
The $1/f$ dependence of the waveform signals a very strong burst\footnote{We can roughly define a burst as 
a wave with a long tail in the frequency domain. The longer the tail in frequency domain the sharper the emission 
was in the time domain. For comparison, bursts from cusps and kinks go like $f^{-4/3}$ and $f^{-5/3}$ respectively \cite{DV1}. The bursts we consider here are much stronger.} and leads to a scale invariant spectrum with constant power emitted per logarithmic frequency interval, as we will now show. 

\subsection{Energy and Power}
The power radiated per frequency interval is \cite{MV}
\be
P_n = 2 n^2 \pi^2 G\mu^2 \int_0^1du\left(1/u -u\right)^2|I_n(u)|^2\,\,,\label{eq:powern}
\ee
where $I_n(u)$ is given by Eqn.(\ref{In}).
Again, there are two interesting frequency intervals, the `scale
invariant' range $1 \lesssim n \ll \gamma_0^2$, and the `convergent
range' $1 \ll \gamma_0^2 \ll n$.
We can analytically determine the power radiated in each mode in the limit $\gamma_0 \to \infty$:
\be
P_n = \frac{4 G\mu^2}{n^2\pi^2}\left(3 \gamma_E - 4 \mbox{Ci}(2\pi n) + \mbox{Ci} (4\pi n) + \log(4 n^3 \pi^3) + 4 n\pi \mbox{Si} (2\pi n) - 2 n\pi \mbox{Si} (4 \pi n)\right)\,\,,
\ee
where Ci and Si are the cosine integral and sine integral functions, respectively, and $\gamma_E$ is the Euler constant.
We can find the beaming angle by examining the integrand in Eq.
(\ref{eq:powern}).  In the `scale invariant' range, the radiation is
beamed in a cone of apex angle $\theta_n \approx \sqrt{1/n}$, and the
power in the $n^{th}$ mode is approximated by
\be
P_n \approx \frac{4 G\mu^2}{n}\,\,.
\ee
The `scale invariant' part of the radiation is thus constant per
logarithmic frequency interval, 
and it dominates the total power
radiated, 
\begin{equation}
\label{power0}
P \approx 4 G\mu^2 \sum_{n =1}^{\gamma_0^2} \frac{1}{n } \approx 4 \ln(\gamma_0^2) G\mu^2\; .
\end{equation} 
There is a small correction to this formula coming from the $1/n^2$ part at
high frequency but it can be safely neglected for large enough
$\gamma_0$. It is instructive to re-derive the waveform from energy considerations. For a single burst, the
meaningful quantity is the energy
\be
dE \approx \frac{4 G\mu^2ldf}{f} \qquad \quad 1/(1+z)l\, \leq \,f\, \leq \,\gamma_0^2/(1+z)l\,\,,\label{eq:burstenergy}
\ee
which is beamed in a cone of angle $\theta_f \approx
\sqrt{1/(1+z)fl}$. The gravitational wave energy from an isotropic burst is
given by
\be
dE = 4\pi r^2 M_P^2 f^2 h^2(f) df\,\,.
\ee 
Thus if all the radiation is contained in a cone of solid angle
$\theta_f^2$, the waveform must satisfy
\be
dE =  r^2 \theta_f^2 M_P^2 f^2 h^2(f) df\,\,,
\ee
and so comparing with Eq.(\ref{eq:burstenergy}) yields the waveform
(assuming the burst is aimed directly at us)
\be
h(f) \sim \frac{G\mu}{r}\frac{l}{f}\,\,.
\ee
For observation within a narrow frequency band, the beaming angle {\em appears} narrowed by
redshift.  This is because an observation of a frequency $f$ today
corresponds to a higher emitted frequency, and thus a narrower cone of radiation.  Since the energy in a gravitational burst must scale like $1/a$, all of the redshift dependence is in
the observed beaming angle, and the waveform remains unchanged.  This means that sources at very large redshift are just as bright but are less often directed at us.

\subsection{Beyond the Straight String Approximation}

In the idealized straight string approximation, at the time of fly-by all the energy of the
string/beads system is located in a region smaller than the corresponding
Schwarzschild radius, and a black hole should form.  
This will not occur for segments with some minimum impact parameter, due to angular momentum.  Given that the
Schwarzschild radius is $r_s = \frac{M_{BH}}{m_p^2}$, and that the
energy of our system is of the order $\mu l$, we will form a black whenever the impact parameter $b$ satisfies
\be\label{bhbound}
b < \frac{\mu l}{m_p^2} \sim G\mu l\; .
\ee
In this paper we work under the assumption that for most strings, the
impact parameter is always larger than this bound.  This is legitimate,
since for a network of scaling cosmic strings
the segments will contain a number of kinks.
Kink-bead interactions are very similar to fly-bys, since in both cases the bead's acceleration
is piecewise constant with sharp discontinuities.
Such strings will not form black holes since the bead impact parameter is of order the length of the segment.
If the average number of kinks per segment falls significantly below one, black hole formation must be
considered.  This may yield interesting phenomenology in its own right.

One may wonder whether the presence of kinks will alter the observable signal from such a network 
and here we argue that it does not.  We will assume that
the kink angle is quite small, in which case it may be thought of as a
perturbation on a straight string.  

Under this assumption, we expect a string with $k$ weak kinks to produce the same
total power, but divided into $k$ separate bursts.  Starting fully
extended and at rest, the beads will accelerate toward the segment
center of mass, and encounter $k$ kinks on the way.  The first
kink will cause a burst in the low frequency range, since the bead has
not had a chance to reach its top speed.  Successive bursts will be
higher and higher in frequency, since the bead is moving faster and
faster, until the maximum speed is reached. 

Since the kinks have little effect on the bead velocity we can simply
divide up the (flat) energy spectrum of the straight string scenario
into $k$ distinct domains.  This means that the each of the $k$ bursts
represents a different interval in frequency space, but their
amplitude is identical to the simple case.  In essence, the kinks act like
a prism, dividing the burst into coequal frequency domains.  For experiments sensitive to either a narrow frequency
band or total power, there can be no distinction between a single co-linear bursts, and prismatically scattered bursts.

\subsection{The Rate of Bursts}

Let the number density of bead-antibead pairs connected
by a string of length between $l$ and $l+dl$ at time $t$ be
\begin{equation}
\label{nofladnt}
n(l,t)dl.
\end{equation}
By length here we mean the fully stretched out length of the system
when the beads are at rest.  Each bead-antibead pair oscillates in a
time $T \sim l$ and the number of bursts per unit spacetime volume we
observe at time $t$ is
\begin{equation}
\label{nuofladnt}
\nu(l,t)dl \sim \frac{1}{l} n(l,t)dl.
\end{equation}
We have restricted ourselves to the case when the beads are
relativistic and the radiation is highly beamed. We do not observe all
bursts, rather we only observe a fraction
\begin{eqnarray}
\label{Delta}
\Delta(l,f,z) \sim
\frac{1}{4 (1+z) f l} \Theta(\gamma_0 - 1) \Theta\left((1+z)fl - 1
\right) \Theta\left(\gamma_0^2 - (1+z)fl)\right) , 
\end{eqnarray}
with $\gamma_0 \sim \mu l/m$.  The first term on the right hand side
is the beaming fraction for an angle $\theta_f \sim ((1+z)fl)^{-1/2}$,
the first Heaviside $\Theta$-function (second term) ensures that we
are in the relativistic regime $\gamma_0 \gg 1$, and the second and
third Heaviside $\Theta$-functions ensure that we are in the regime
given by Eq.~(\ref{freqrange}) where the waveform
Eq.~(\ref{finvwaveform}) is valid.  Although in Eq.~(\ref{Delta}) we
have included the upper bound on the frequency for clarity, it turns
out to be irrelevant. The condition $\gamma_0^2 > (1+z)fl$ can be turned
into a bound on the length of the segment of string connecting the beads,
$l > (1+z) f (G\mu)^{-1} \kappa l_p^2$, where $l_p$ is the Planck
length. For typical values of $G\mu$ and $\kappa$ even for the largest
possible redshifts the lower bound on $l$ is within a few orders of magnitude
of the Planck length.

Time and distances can be written as cosmology dependent functions of the
redshift $t(z)=H_0^{-1}\varphi_t(z)$, $r(z) = H_0^{-1}\varphi_r(z)$
and $dV(z)=H_0^{-3}\varphi_V(z)dz$. 
These functions depend on cosmological parameters and need to be
computed numerically as described in Appendix A of~\cite{SCMMCR},
where a vanilla $\Lambda$-CDM model was used with recently measured
parameters that can be found in~\cite{hubble}.  One can obtain
reasonable (good to about 20\%) analytic approximations for the cosmological
functions using variations of the expressions introduced
in~\cite{DV1}.  In particular the following expressions,
\begin{eqnarray}
\varphi_t &= &(1+z)^{-3/2} (1+z/z_{eq})^{-1/2},\\
\label{eq:phit}
\varphi_r &=& \frac{z}{1+z/3.5042},\\
\label{eq:phir}
\varphi_V&=&12z^2 (1+z/1.6)^{-13/2} (1+z/z_{eq})^{-1/2},
\label{eq:phiV}
\end{eqnarray}
seem to work well (with $z_{eq}=5440$).  We note that for consistency, the value of $H_0$
that should be used with these functions is $73$
km/s/Mpc~\cite{hubble}.  This allows us to
write the rate of bursts we expect to see from a volume $dV(z)$ (the
proper volume in the redshift interval $dz$), from
bead-antibead pairs with lengths in the interval $dl$ as,
\begin{eqnarray}
\label{dRdzdl}
\frac{dR}{dV(z)dl} &= &(1+z)^{-1} \nu(l,z) \Delta(l,f,z)\; ,\nonumber \\
\frac{dR}{dzdl} &=& H_0^{-3}(1+z)^{-1} \varphi_V(z) \nu(l,z) \Delta(l,f,z).
\end{eqnarray}
The first $(1+z)$ factor  comes from the relationship between the observed burst
rate and the cosmic time.  To compute the rate of bursts an estimate of the minimum detectable
amplitude $A_{\rm min}$ needs to be performed (see section~\ref{AMIN}).  A matched
filter search returns the amplitude $A$ of a template,
\begin{equation}
\label{matchedfilter}
h(f)=At(f)
\end{equation}
where in this case $t(f)=f^{-1}$, and the dimensionless amplitude is given by
\begin{equation}
\label{amplitude}
A = C\frac{G\mu l H_0}{\varphi_r(z)}\,\,,
\end{equation}
where again $C = \sqrt{8}\sin^2(\sqrt{2})/\pi$.  We can write the length as a function of the amplitude,
\begin{align}
\label{length3}
l(A,z) &=  \frac{A \varphi_r(z)}{C G \mu H_0}, & \frac{dl}{dA} = \frac{l(z,A)}{A},
\end{align}
and change variables to write the rate as a function of the amplitude.
In the end we perform the integrals to compute the rate above the
minimum detectable amplitude
\begin{equation}
\label{R}
R_{>A_{\rm min}}=\int_{A_{\rm min}} ^{\infty} dA \int_0^{\infty} dz\frac{dR}{dzdA}
\end{equation}
and see how that rate compares with one event per year (say).  
In the following sections, we will compute $n(l,t)$ and $A_{min}$ and the results for the rate is shown for LIGO, Advanced LIGO, and LISA in Fig.(\ref{burst}).

\subsection{Stochastic Background}

A stochastic background of gravitational waves is produced by the
incoherent superposition of all bursts ever produced by the network.
It is typically expressed as the ratio of the energy density of
gravitational waves to the critical density today. To derive an expression for the stochastic background 
we begin with the expression for the plane wave expansion for the
metric perturbation \cite{allenandromano},
\begin{equation}
h_{ab}(t,\vec x)=\sum_A\int_{-\infty}^\infty df \int_{S^2} 
d\hat\Omega\ h_A(f,\hat \Omega) e^{i2\pi f(t-\hat \Omega\cdot\vec x/c)}\ 
e_{ab}^A(\hat \Omega) .
\label{e:h_ab}
\end{equation}
Here $A=+,\times$ labels polarisations, ${\hat \Omega}$ is the
direction of propagation of the gravitational wave and $e_{ab}^A(\hat
\Omega)$ is the polarisation tensor.
If the stochastic background is isotropic, stationary, unpolarised, and
uncorrelated the ensemble average of the Fourier amplitudes
$h_A(f,\hat\Omega)$ satisfies \cite{allenandromano}
\begin{equation}
\langle h_A^*(f, \hat \Omega) h_{A'}(f',\hat \Omega')\rangle=
\delta^2(\hat \Omega ,\hat \Omega')\delta_{AA'}\delta(f-f')\ H(f).
\label{e:ev1}
\end{equation}
This expression {\it defines} $H(f)$.  The relation between $H(f)$ and $\Omega_{\rm gw} (f)$ is
\cite{allenandromano}
\begin{equation}
H(f)={3H_0^2\over 32\pi^3}\ |f|^{-3}\ \Omega_{\rm gw}(|f|)\ .
\label{e:H(f)}
\end{equation}
where $H_0$ is the Hubble parameter which we take to be $H_0 = 73$
km/s/Mpc \cite{hubble}.  In the following we describe how to compute $H(f)$ using
Eq.~(\ref{e:ev1}).  We start with an expression for the metric
perturbation as a function of the frequency evaluated at a particular
point in space (we take $\vec x = \vec 0$ for convenience),
\begin{equation}
h_{ab}(f,\vec 0)=\sum_A \int_{S^2} 
d\hat\Omega\ h_A(f,\hat \Omega) e_{ab}^A(\hat \Omega) .
\label{e:h_aboffat0}
\end{equation}
From Eq.~(\ref{e:h_aboffat0}) it is clear that if we multiply both
sides of Eq.~(\ref{e:ev1}) by $e^{ab}_{A'}(\hat \Omega')$, integrate
over the sphere ($\int_{S^2} d\hat \Omega'$) and sum over polarisations
($\sum_{A'}$), we obtain,
\begin{equation}
\langle h_A^*(f, \hat \Omega) h^{ab}(f',\vec 0)\rangle=
e^{ab}_{A}(\hat \Omega) \delta(f-f')\ H(f).
\label{e:ev2}
\end{equation}
Similarly, multiplying both sides by $e_{ab}^{A}(\hat \Omega)$,
integrating over the sphere ($\int_{S^2} d\hat \Omega$) and summing over
polarisations ($\sum_{A}$), we obtain,
\begin{equation}
\langle h_{ab}^*(f, \vec 0) h^{ab}(f',\vec 0)\rangle=
16 \pi \delta(f-f')\ H(f).
\label{e:ev3}
\end{equation}
Here we have used the fact that $\sum_{A} e_{ab}^{A} e^{ab}_{A} =4$. 
To remove the delta function we borrow a trick from \cite{DV1}.  In
the limit $f \rightarrow f'$ the delta function takes the formally
infinite time interval $\delta(f-f') \rightarrow T$, so that  
\begin{equation}
\langle |h_{ab}(f, \vec 0)|^2\rangle = 16 \pi T \ H(f).
\label{e:ev4}
\end{equation}
The factor of $T$ will cancel eventually.  We have seen that the gravitational wave signal is linearly polarised which means that
for a wave traveling along the $\hat z$ direction we can write the
metric perturbation as
\begin{equation}
h_{ab}(f, \vec 0)=h(f)\left(
\begin{array}{ccc}
\cos 2\psi &  \sin 2\psi & 0 \\
\sin 2\psi & -\cos 2\psi & 0 \\
0 & 0 & 0\\
\end{array}
\right).
\end{equation}
If we average over the polarisation angle $\psi$ we get $\langle
h^*_{ab}(f, \vec 0)h^{ab}(f, \vec 0)\rangle = 2 \langle
|h(f)|^2\rangle$, and hence,
\begin{equation}
H(f) = \frac{1}{8 \pi T} \langle |h(f)|^2\rangle.
\label{e:ev5}
\end{equation}
Thus we can write,
\begin{equation}
\Omega_{\rm gw}(|f|) = \frac{4 \pi^2}{3H^2_0}f^3 \frac{1}{T}\langle |h(f)|^2\rangle.
\label{e:Omega(f)}
\end{equation}
The strain (Eq.~\ref{finvwaveform}) and the rate of bursts (Eq.~\ref{dRdzdl}) enter the ensemble averaged strain through
\begin{equation}
\langle |h(f)|^2\rangle = T \int dz \int dl \, h^2(f,z,l)  \frac {dR}{dzdl}.
\label{e:evh(f)sq}
\end{equation}
Thus, the stochastic background produced by the incoherent
superposition of bursts is
\begin{equation}
\Omega_{\rm gw}(f) = \frac{4 \pi^2}{3H^2_0}f^3 
\int dz \int dl \, h^2(f,z,l)  \frac {dR}{dzdl},
\label{e:Omega(f)2}
\end{equation}
where the integrals are over all redshifts $z$ and over all lengths
$l$ connecting the bead pairs.  

 As pointed out by Damour and Vilenkin~\cite{DV1}, if we are interested
in the confusion noise produced by the network of beads and strings we
must avoid the biasing of $\Omega_{\rm gw}(f)$ that could result from
including large amplitude rare events.  We can accomplish this by excluding events
that occur less frequently than the typical duration of bursts.  The
duration of a burst as seen in an instrument with peak sensitivity at
frequency $f$ is ${\cal O}(f^{-1})$.  Only bursts that overlap, {\em i.e.,}
small amplitude frequent bursts contribute to the Gaussian background
we are trying to estimate. We can follow the procedure developed
in~\cite{SMC}.  We begin by expressing the rate as a function of the
strain and the redshift rather than the length and the redshift.  We
can do this change of variables using Eq.~(\ref{strain3cont}) to write
\be
\frac{dR}{dhdz}=\frac{l}{h}\frac{dR}{dldz}.
\ee
We then integrate the rate over all redshifts and find the strain $h_*$ 
for which the rate at and above that strain becomes smaller than the peak 
experiment frequency $f$.  Namely we find the value of $h_*$ such that, 
\be
R(>h_*)=\int_{h_*}^\infty dh \int_0^{\infty} dz \frac{dR}{dhdz},
\ee
and evaluate
\begin{equation}
\Omega_{\rm gw}(f) = \frac{4 \pi^2}{3H^2_0}f^3
\int_0^{h_*} dh h^2 \int_0^\infty dz \,  \frac {dR}{dzdh}.
\label{e:Omega(f)3}
\end{equation}
Equation~(\ref{e:Omega(f)3}) gives the so-called confusion noise
produced by the network, a Gaussian background that
needs to be included along with other sources of noise to estimate
the total noise present in our detectors. If the confusion noise is
sufficiently large it can have adverse effects on our ability to
detect individual bursts.

For stochastic background searches that use cross correlations between
detectors, such as those performed by LIGO, it is not necessary to
remove the large amplitude events in our estimate of $\Omega_{\rm
  gw}(f)$, and Eq.~(\ref{e:Omega(f)2}) may still be used.  Large
infrequent bursts will contribute to the cross-correlation in our
estimate of $\Omega_{\rm gw}(f)$, even though cross-correlations are
not the optimal procedures for finding such bursts.  These large
infrequent bursts are optimally detected using other techniques:
Matched-filters if they are sufficiently large amplitude (as described
above), or methods optimized for non-Gaussian stochastic background
detection~\cite{Drasco:2002yd}.

\section{Cosmological Evolution}
\label{cosmology}

The cosmic string network forms at the end of brane inflation (or
similar hybrid inflationary scenario) without any beads present.  This
infinite string network begins to decay immediately, resulting almost
exclusively in super-horizon length segments.  As long as the typical
segment length is super-horizon, the evolution will be identical to
that of an infinite string network, and scaling is readily expected
despite the presence of a small number of sub-horizon (oscillating)
segments.

At a time $t_*$ the typical segment length becomes sub-horizon.  We
will call this the beginning of the short string epoch.  Much like matter-era cosmic
strings, these segments are smoother and straighter than the segments
during the long string epoch.  The strings during this epoch will
cease to obey a scaling solution.  This is because the segments will have
a drastically reduced intercommutation probability due to low peculiar
velocities.  The segment energy density will thus increase relative to the
scaling solution during the short string epoch until either it begins to lose significant energy to gravitational radiation, or the universe becomes matter dominated.  The violation of scaling during the short string epoch can thus increase the energy density relative to long cosmic strings by a factor of at most $(G\mu)^{-1/4}$, which is
significant in our parameter space.  The network will continue to
fragment into shorter segments, causing the energy loss to
gravitational radiation to eliminate most of the network around the time $t =
t_{**}$.  
The primary phenomenological quantity that we must determine is the number
density of string segments of a given energy.

\subsection{Number Density of Segments: $n(l,t)dl$}
Let $n(l,t)dl$ be the number density of bead-antibead pairs connected by a string of
length between $l$ and $l+dl$ at time $t$.  As before, we define the length $l$
of a segment by its maximum possible length: $l = E/\mu$.  Then the
energy density in cosmic string is given by
\be \label{segmentsum}
\int_0^\infty  n(l,t)\mu ldl = \rho_{cs}. 
\ee
Notice we do not need to distinguish between ``infinite strings" and finite segments:  
the former are more accurately characterized as having a super-horizon length.  We do
not track loops explicitly.  Instead, we note that loops are not produced by segments which
have begun to oscillate.
Before we perform a careful derivation of $n(l,t)dl$, it is enlightening to first
consider a simple estimate that can be done in the scaling regime.
We can treat the
cosmic string network as a single worldsheet where decay events are
uncorrelated points on the sheet.  This means a metastable cosmic
string of decay rate per unit length $\Gamma_2$ will have decay events
obeying a Poissionian distribution.  The probability of a worldsheet
proper area $A$ containing $k$ nucleation events is given by
\be \label{poissonian}
P_k(A) = \frac{(A\Gamma_2)^k e^{-A \Gamma_2}}{k!}\,\,.
\ee
Thus at a time $t$, the probability for a given string to be of length $l$
is given by the probability that exactly zero decay events have
occurred in the worldsheet area $lt$, {\em i.e., in the entire past of
  the segment}, times the probability that one or more decay events
did occur in an area $t dl$.  (We are asking for a segment of length
at least $l$ and at most $l + dl$.)  Using Eq.~(\ref{poissonian}) the
probability of a segment having such a length is given by
\be\label{lengthdistribution}
P(l,t)dl = \Gamma_2  t \exp (-\Gamma_2 l t ) dl\,\,.
\ee
We can relate $P(l,t)dl$ to the number density $n(l,t)dl$ via Eq.~(\ref{segmentsum})
\be
n(l,t)dl = P(l,t)dl\frac{\rho_{cs}(t)}{\int_0^\infty P(l',t)\mu l' dl'}
\ee
where $\rho_{cs}(t)\sim \mu/t^2$ is fixed by the scaling solution.  Thus we have
\be \label{nsimple}
n(l,t)dl =  \Gamma_2^2\exp\left(-l t \Gamma_2\right)dl
\ee
This simple 
argument is expected to be valid in the scaling regime when most of the strings are super-Hubble in length.  The number density of segments of finite length goes to zero both when $\Gamma_2 \rightarrow 0$ (the network is stable) and when $\Gamma_2 \rightarrow \infty$ (the network decays immediately).  For a fixed length and time, $n(l,t)$ has a maximum at $\Gamma_2 = 1/lt$.  Therefore the time ($t_*$) at which a significant fraction of the segments in the network are subhorizon $l \sim t$ is
\be
\left<l\right> \sim H^{-1} \quad \Longleftrightarrow \quad t = t_* := \Gamma_2^{-1/2}
\ee
To correctly account for the deviation from scaling and the gravitational back-reaction, we need a more thorough analysis.  Because $n(l,t)$ is a length density it satisfies the continuity equation 
\begin{equation}
\label{eq:boltzmann}
\frac{\partial n(l,t)}{\partial t} = -\frac{\partial }{\partial
  l}\left(\dot l n(l,t)\right)- 3\frac{\dot a}{a} n(l,t) + g\,\,,
\end{equation} 
where $gdldt$ is the number density of pairs
produced with lengths between $l$ and $l+dl$, at a time between $t$
and $t+dt$.  The function $\dot{l}$ incorporates the effects of the geometry on individual segments, {\em i.e.,} Hubble stretching and drag.  We can characterize explicit terms within $g$ as being one of three
types:  loop producing, segment intercommutation, and segment breaking.  Loop production
provides a stable solution for $\rho_{cs}(t) = \mu \int n(l,t)ldl$, the scaling solution.  Two-segment intercommutation will not be explicitly important and breaking will eventually destroy the scaling solution.  We write 
\be
g = g_{\tt loop} + g_{\tt ic}+ g_{\tt break}\,.
\ee
In Appendix \ref{sec:distribution}, we discuss each term in more detail.  Here we simply argue that  intercommutation and loop production ensure scaling in the absence of breaking.\footnote{The main effect of intercommutation is to create a small scale structure on the strings that leads to the creation of loops.  Loop production is the dominant channel for energy losses in the network.  It is also sufficient for scaling - one can turn off gravity and the ``infinite" network still scales.}  
To make the problem tractable, we will coarse grain over the scale of loop production and treat it as a continuous shortening
of segments.  The result is an $\dot l$ which now includes the effects
of both the geometry (expansion) and loop production, 
\ba
\dot{l}_{\tt} &=& \dot{l}_H + \dot{l}_{\tt loop}\\
&=& 3 H l - \frac{2 l}{t}\quad\quad{\mbox{(scaling solution)}}
\ea
It is readily seen that the Boltzmann equation (Eq.~(\ref{eq:boltzmann})) with $g=0$ and the above $\dot l$ has a scaling solution with $\rho_{cs} \sim \mu/t^2$ in both the matter and radiation eras.  We are now ready to include breakage with 
\ba\label{gbreak}
\hspace{-3mm}g_{\tt break}dl &=&  \Gamma_2\left(2\int_l^\infty \hspace{-1mm}n(l',t)dl' \,\, - ln(l,t)\right)dl
\ea
The breakage formula can be motivated as follows.
The breakage rate of strings of length $l$ is $\Gamma_2l$, which explains the last term.
The first term can be understood by considering the process by which $n(l,t)dl$ increases.  This
is entirely due to breaking longer strings, those of length $l' > l$.  The rate at which longer strings break to produce those of length between $l$ and $l + dl$ is given by the number of longer strings present (hence the integral), and the measure of string where a break yields a shorter string of length between $l$ and $l + dl$, {\em i.e.,} $2dl$.  The factor of two is because the breaking point can be closer to either of the two endpoints of the longer string.  Because string breakage should not affect the string energy density, 
we can easily check that $\frac{d}{dt} \int_0^\infty \mu l n(l,t)dl$ is independent of $\Gamma_2$, which it is.  Finally, the last ingredient we will need to solve the Bolztmann equation is the energy loss in gravitational waves as sub-horizon strings oscillate.  We already calculated the total power radiated by an oscillating segment\cite{MV},
\begin{equation}
\label{power}
P \approx 4 \ln(\gamma_0^2) G\mu^2,
\end{equation} 
and the energy comes out of the length of string connecting the pair
(we think of the beads coming to rest, and with each oscillation
the string gets a little shorter because of GW emission).  Thus the
length of string (when the beads are at rest) decreases as
\begin{equation}
\label{power2}
\dot l_{\tt gw} \approx - 4 \ln(\gamma_0^2) G\mu \qquad l \lesssim  H^{-1}
\end{equation} 
We will parametrize this as $\dot{l}_{\tt gw} = - \Gamma G\mu$ with typical values $\Gamma \sim 50$ due to large $\gamma$ factor of the beads.
The phase transition at $t = t_*$ leads to a network that
consists of subhorizon segments, which are straightened by expansion, but not stretched.   These will oscillate and fragment until their GW radiation becomes dominant, at a time
of order $t_{**}$, after which the network disappears.  We can estimate $t_{**}$ by comparing the
power lost to scaling versus the power lost to gravitational radiation:
\ba
\frac{d}{dt} \left(\frac{\mu}{t^2}\right) &\sim& \Gamma G\mu^2 \int_0^\infty \hspace{-1mm}n(l,t)dl\\ \Rightarrow \,\,t_{**} &=&  t_*/\sqrt{\Gamma G\mu}\,\,=\,\,1/\sqrt{\Gamma_2\Gamma G\mu}
\ea

Notice $t_{**}$ does not depend on expansion, unlike $t_*$.  We should therefore think of $t_{**}$ as the network lifetime irrespective of the scaling solution\footnote{We thank K. Olum for pointing this out.}.
\subsection{Solutions}
During the long string epoch, the network will satisfy the scaling Boltzmann equation (now an integro-differential equation given $g_{break}$ in Eq.~\ref{gbreak}),
\begin{equation}
\label{eq:longboltzmann}
\frac{\partial n(l,t)}{\partial t} = -\frac{\partial }{\partial
  l}\left[\left(3 H l - \frac{2 l}{t}\right) n(l,t)\right]- 3H n(l,t) + g_{\tt break}\,\,,
\end{equation} 
whose solution is
\be
n(l,t) =  4\Gamma_2^2\exp(-2\Gamma_2lt)\,\,,
\ee
where we have assumed radiation dominance for now.  This solution is, up to a factor of two in the exponential, exactly what we obtained in our simplistic derivation of Eq. (\ref{nsimple}).  After the long string epoch, the strings will begin to oscillate, which will drastically alter their appearance and behavior. 
In particular, we can now ignore Hubble stretching and loop production, and therefore the only
contribution to $\dot l$ is from energy loss to gravitational radiation.  The Boltzmann equation then reads
\begin{equation}
\label{eq:shortboltzmann}
\frac{\partial n(l,t)}{\partial t} = -\frac{\partial }{\partial
  l}\Gamma G\mu \,n(l,t)- 3H n(l,t) + g_{\tt break}\,\,,
\end{equation} 
which has the solution
\be\label{nradbetter}
n(l,t) = \Gamma_2^2\exp(-\Gamma_2 l t)\sqrt{\frac{t}{t_*}}\exp(-\tfrac{1}{2}\Gamma G\mu \Gamma_2 t^2)\,\,.
\ee
where one recognizes the previously defined $t_{**} = 1/\sqrt{\Gamma G\mu \Gamma_2}$ in the exponential.  The complete evolution of the segments (including the cases where the network decay in the matter era) is extremely well approximated by
\be \label{nbetterB}
n(l,t) =  \Gamma_2^2 \sqrt{\frac{{\mbox {med}}\left\{t,t_*,\max\{t_*,t_{eq}\}\right\}}{t_*}} e^{-\Gamma_2l(t + \min\{t,t_*,t_{eq}\})} e^{- \tfrac{1}{2}t^2/t_{**}^2}\,\,,
\ee
where ${\mbox{med}}\{\}$ is the median.  The exact solution is given
in Appendix \ref{sec:distribution}.

\section{Results}
\label{results}

In this section we examine the parameter space of the theory to find 
detectable bursts and stochastic background signals.

\subsection{The Minimum Detectable Burst Amplitude}
\label{AMIN}

To produce an estimate of the minimum detectable amplitude we follow
the analysis in~\cite{SCMMCR} for cosmic string cusps closely.  As we
have seen, the gravitational waveforms in the frequency domain are
\begin{equation}
h(f) = A t(f),
\label{eq:hf1}
\end{equation}
with $t(f)=f^{-1}$.  We will use the conventional
detector-noise-weighted inner product \cite{cutlerflanagan},
\begin{eqnarray}
(x|y) \equiv 4 \Re \int_0^\infty df \,\frac{x(f)y^*(f)}{S_h(f)},
\label{eq:dorprod1}
\end{eqnarray}
where $S_h(f)$ is the single-sided spectral density defined by
$\langle n(f) n^*(f')\rangle = \tfrac{1}{2} \delta(f-f') S_h(f)$,
where $n(f)$ is the Fourier transform of the detector noise.  The
template $t(f)$ can be normalised using the inner product
$\sigma^2=(t|t)$ so that ${\hat t}=t/\sigma$, and $({\hat t}|{\hat t})=1.$
For an instrumental output $s(t)$, the signal to noise ratio
(SNR) is defined as,
\begin{equation}
\rho \equiv (s|{\hat t}).
\label{eq:snr1}
\end{equation}
In general, the instrument output is a burst $h(t)$ plus some noise
$n(t)$, $s(t)=h(t)+n(t)$.  When the signal is absent, $h(t)=0$, it is
easy to show the SNR
is Gaussian distributed with zero mean and unit variance.
When a signal is present, the average SNR is
\begin{eqnarray}
\langle \rho \rangle &=& \langle (h|{\hat t}) \rangle + 
\langle (n|{\hat t}) \rangle
= (A\sigma {\hat t}|{\hat t} ) = A\sigma,
\label{eq:snr7}
\end{eqnarray}
and the fluctuations also have unit variance.
This means that for an SNR threshold $\rho_{\text{th}}$, on average, 
only events with amplitude
\begin{equation}
A_{\text{th}} \ge \frac{\rho_{\text{th}}}{\sigma},
\label{eq:amp2}
\end{equation}
will be detected.  The form of the Initial LIGO noise curve is well approximated by~\cite{ligosens},
\begin{eqnarray}
S_h(f) &=& \left[ 1.09 \times 10^{-41}  \left( \frac{30\text{Hz}}{f}
\right)^{28} \right.
\nonumber
\\
&+&1.44 \times 10^{-45} \left( \frac{100\text{Hz}}{f} \right)^{4}
\nonumber
\\
&+&  \left. 1.28 \times 10^{-46} 
\left( 1+  \left( \frac{f}{90\text{Hz}}\right)^2 \right)\right]\text{s}.
\label{eq:LIGOIdesign}
\end{eqnarray}
The first term arises from seismic effects, the second 
from thermal noise in the optics, and the third from photon shot noise. 
A reasonable operating point for the
pipeline involves an SNR threshold of $\rho_{\text{th}}=4$.  The value
of the amplitude is related to $\rho$, the SNR, and $\sigma$, the
template normalisation,
\begin{equation}
\sigma^2 = 4 \Re \int_0^\infty df \,\frac{t(f)^2}{S_h(f)}
= 4 \int_0^\infty  df \,\frac{f^{-2}}{S_h(f)}\; .
\label{eq:sigma10}
\end{equation}
Because of Eq.~(\ref{eq:amp2}) an event with an SNR
close to the threshold, $\rho_* \approx 4$
will on average have an amplitude $A_*
\approx 6 \times 10^{-22}$.  Events on average will not be
optimally oriented so we need to increase the amplitude by $\sqrt{5}$
to account for the averaged antenna beam pattern of the instrument.
So we expect $A_{\rm min} \approx 10^{-21}$ for the Initial LIGO noise curve. 
For Advanced LIGO we expect to do about an order of magnitude better
so we take $A_{\rm min} \approx 10^{-22}$.  For LISA, our estimate of the minimum detectable amplitude is $A_{\rm min}= 4 \times 10^{-21}$.
This estimate was made using a LISA noise curve that includes confusion noise from galactic
white dwarf binaries, assuming that some of the binary signals can be fitted
out\cite{Hughes:2001ya,Barack:2004wc}.

\subsection{Burst Detection by LIGO, Advanced LIGO, and LISA}

\begin{figure}
\centering{\includegraphics[width=2.44in]{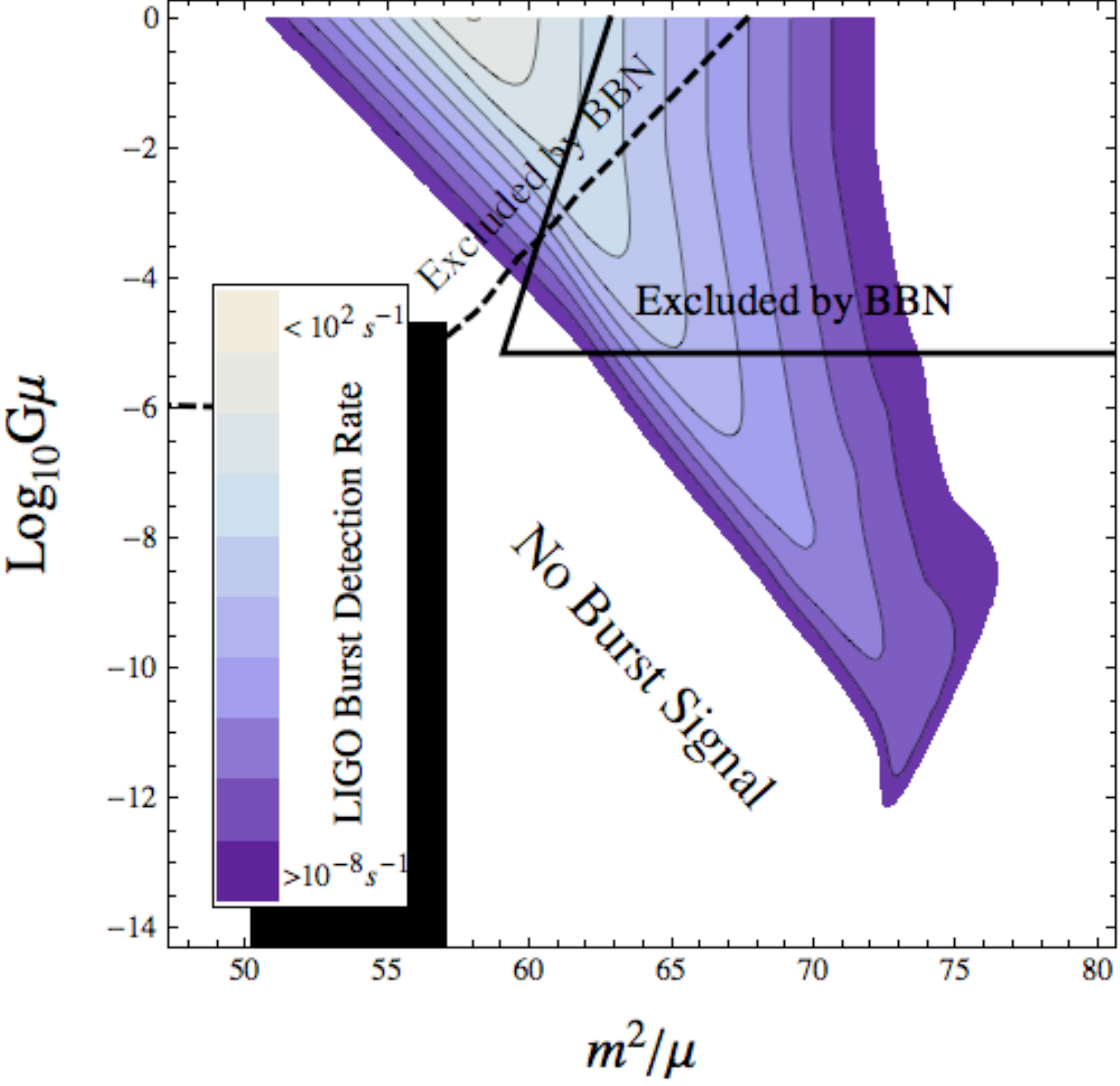}}
\centering{\includegraphics[width=2.22in]{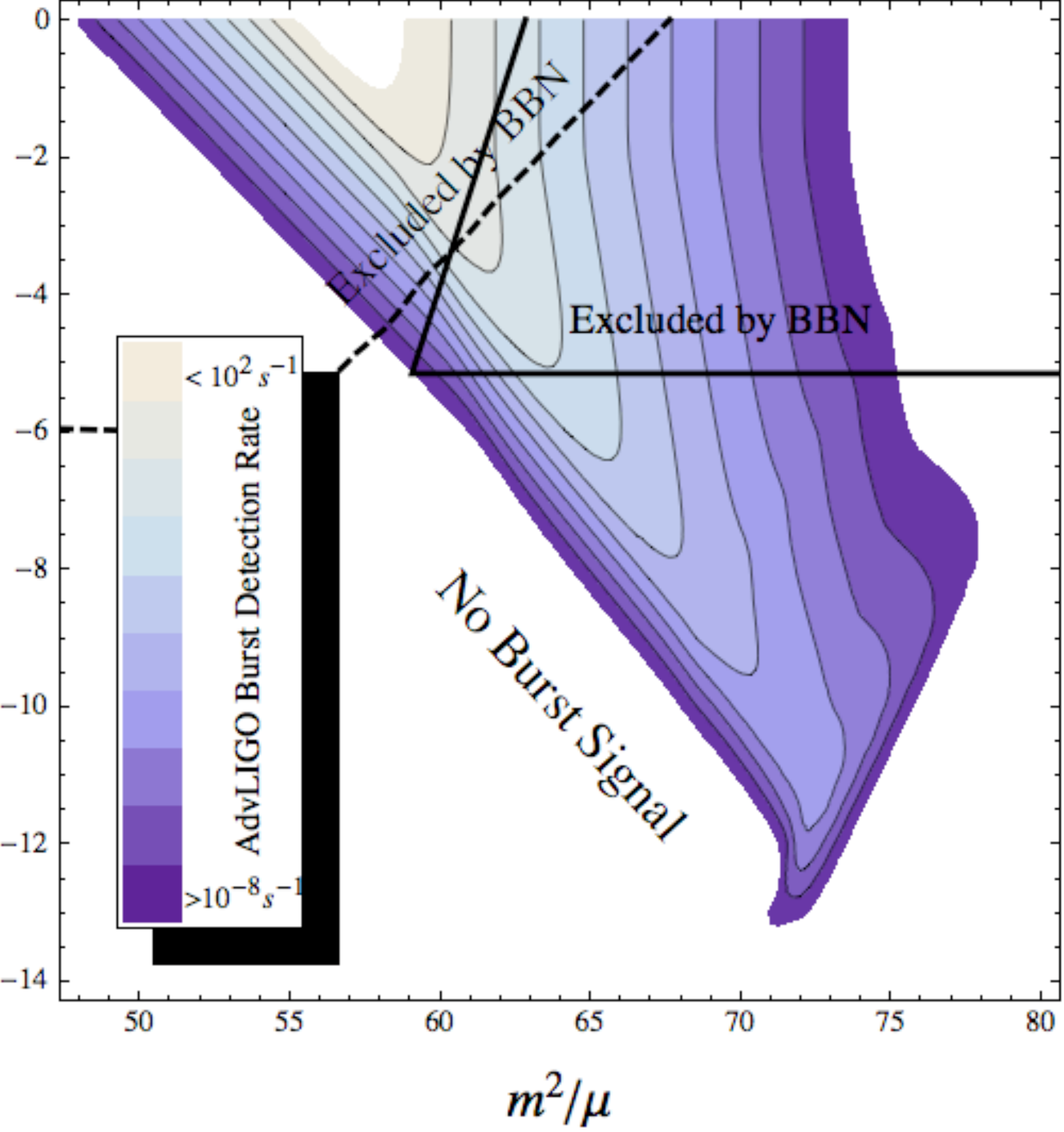}}
\centering{\includegraphics[width=2.22in]{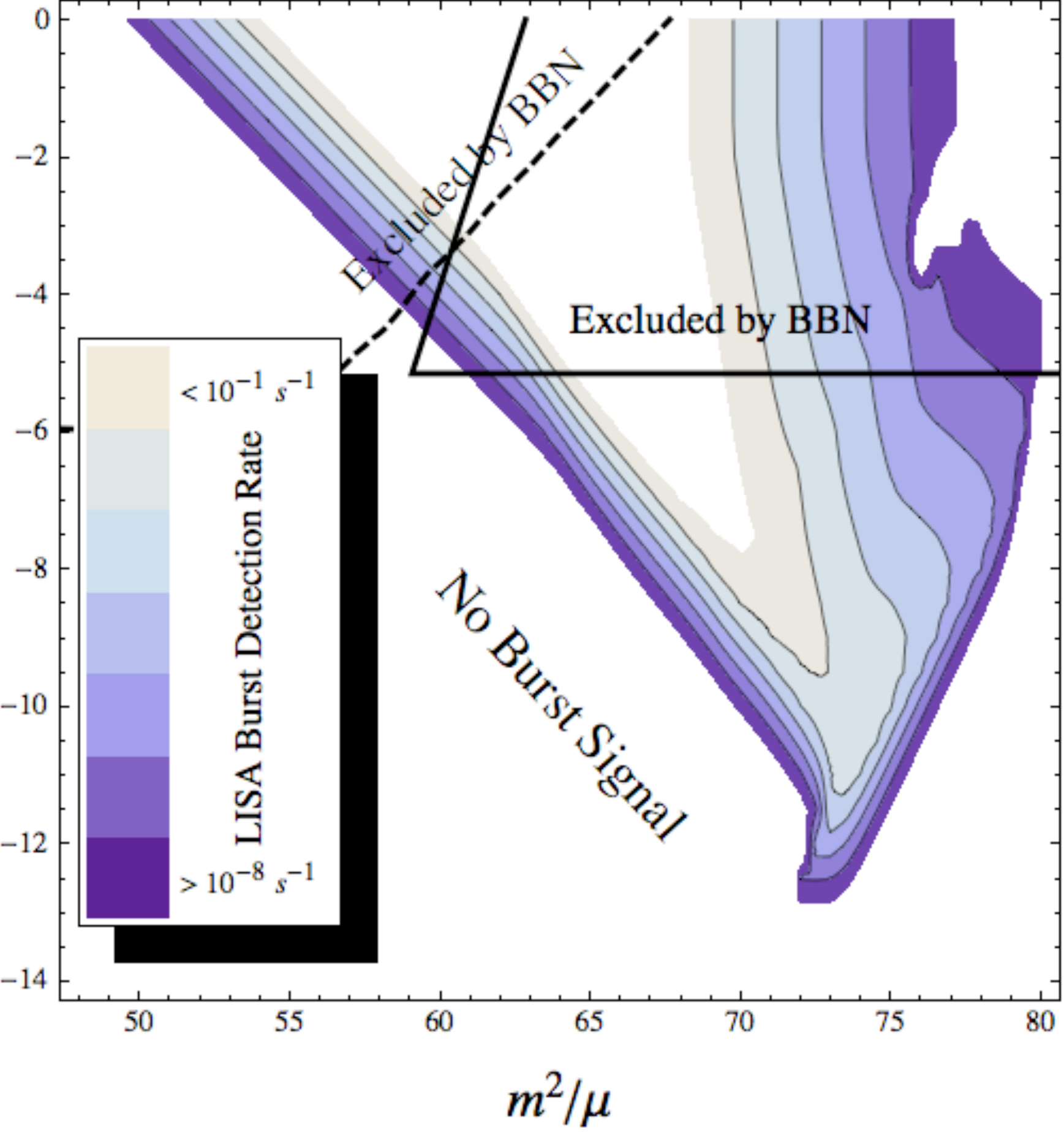}}
\caption{Burst detection rate for LIGO (left), Advanced LIGO (center), and LISA (right).  The solid curve represents
the BBN constraint on string tension from loops {\em i.e.,} long-lived strings, whereas the dashed curve represents the BBN constraint on string tension from bead induced bursts.}
\label{burst}
\end{figure}

We want to calculate the rate of bursts
\begin{equation}
\label{R1}
R_{>A_{\rm min}}=\int_{A_{\rm min}}^{\infty} dA \int_0^{\infty} dz\frac{dR}{dzdA}
\end{equation}
with 
\ba
\frac{dR}{dzdA} & = & \frac{\varphi_V(z) \nu(A, z) \Delta(A,z)}{H_0^3 (1+z)}\frac{m}{2 \mu}\sqrt{\frac{(1+z)\varphi_r(z)}{BG\mu H_0 A} }\,\,,
\ea
which can be obtained from Eq.~(\ref{dRdzdl}) by doing a change of variable from $l$ to $A$ using Eq.~(\ref{length3}).  
The integral over amplitude goes to infinity to good approximation.  More precisely, there is an amplitude (length) cutoff set by the size of the horizon, since only strings with length $l <  1/H$ will oscillate and produce bursts.  

With some approximations, it is possible to obtain an analytic expression for the observed burst rate.  First we will work at large redshift $z \gg z_{eq}$ to simplify all of our cosmological functions. The number density of segments in the radiation era is given by Eq.~(\ref{nradbetter}) and we can make the further approximation that the network decay suddenly at time $t = t_{**}$ and set the density to zero beyond that point.  Therefore we divide the density into the scaling regime $t < t_*$, the non-scaling (short string) regime $t_{*}< t<t_{**}$ and the post string regime $t_{**} < t$.
\ba
n(l,t)  =  4 \Gamma_2^2 e^{-2\Gamma_2 l t} \left\{ \begin{array}{ll} 1  \;\;\; &t\,\, < t_* \\ \frac12 e^{\Gamma_2 lt} \sqrt{t/t_*} \;\;\; &t_{*}<t\,\, < t_{**} \\ 0 \;\;\;  &\qquad  t_{**}<t \end{array} \right.
\ea
With this density, we get 
\ba
\frac{dR}{dzdA} \propto \frac{1}{A^2 f} \left\{ \begin{array}{ll} \frac{(G\mu)^3 M_p^4}{H_0^2 z^7} e^{-2\pi \kappa} e^{-a A/z^2} &  \;\;\; z > z_*\\ \frac{(G\mu)^{13/4}M_p^{9/2}}{H_0^{5/2} z^8} e^{-9\pi \kappa/4} e^{-a A/(2 z^2)} &  \;\;\; z_*>z > z_{**}\\ 0 & \;\;\; z < z_{**} \end{array} \right.
\ea
where 
\be
a \propto \frac{e^{-\pi \kappa} M_p^2}{H_0^2}\; .
\ee
The integrals over $A,z$ can be done exactly, and the results can be written in terms of incomplete gamma functions.  We found that for most parameter values, the detectable signal is dominated by the radiation emitted at $z_{**}$ which can be before or after $z_{eq}$.  We therefore had to resort to numerical integration to incorporate the signal coming from low redshift.  For all numerical work, we assumed the scaling epoch describes an average of 40 long strings per Hubble volume.  The results are shown for LIGO, Advanced LIGO, and LISA in Fig.~(\ref{burst}).  The straight solid lines in all plots delineate the region of parameter space for which radiation from string loops is in conflict with BBN.  In the simplest case, this restricts the string tension to satisfy $G\mu < 7\times 10^{-6}$ \cite{Caldwell:1991jj}.  Because no loops will exist for times after $t_*$, we have shown where this bound can no longer be reliably applied with the vertical kink in the solid curve.  Interestingly, when loops dissappear before BBN, the tension remains constrained by the stochastic radiation from beads.

\subsection{Stochastic Background Observation and Constraints}

\begin{figure}[h]
\centering{\includegraphics[width=5in]{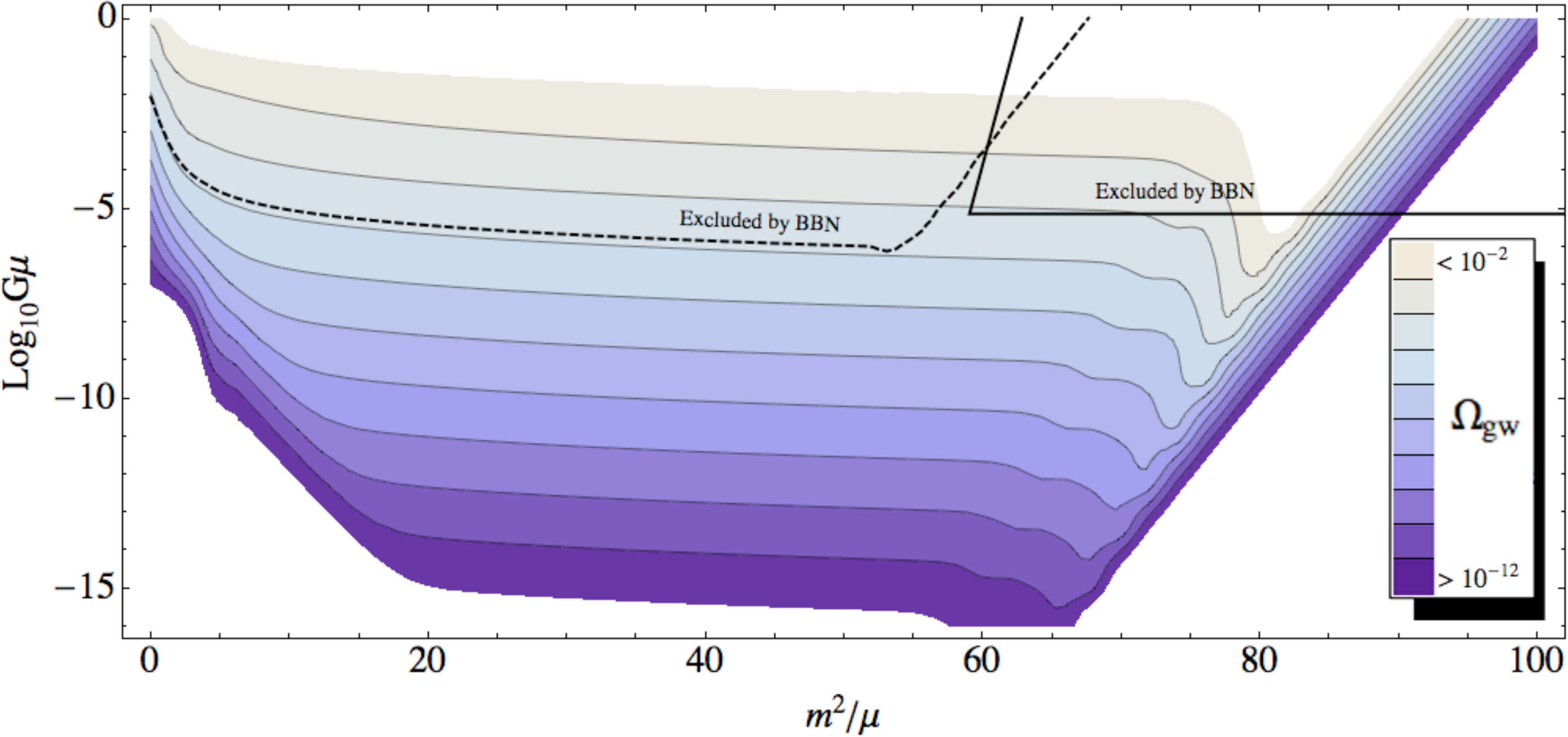}}
\centering{\includegraphics[width=5in]{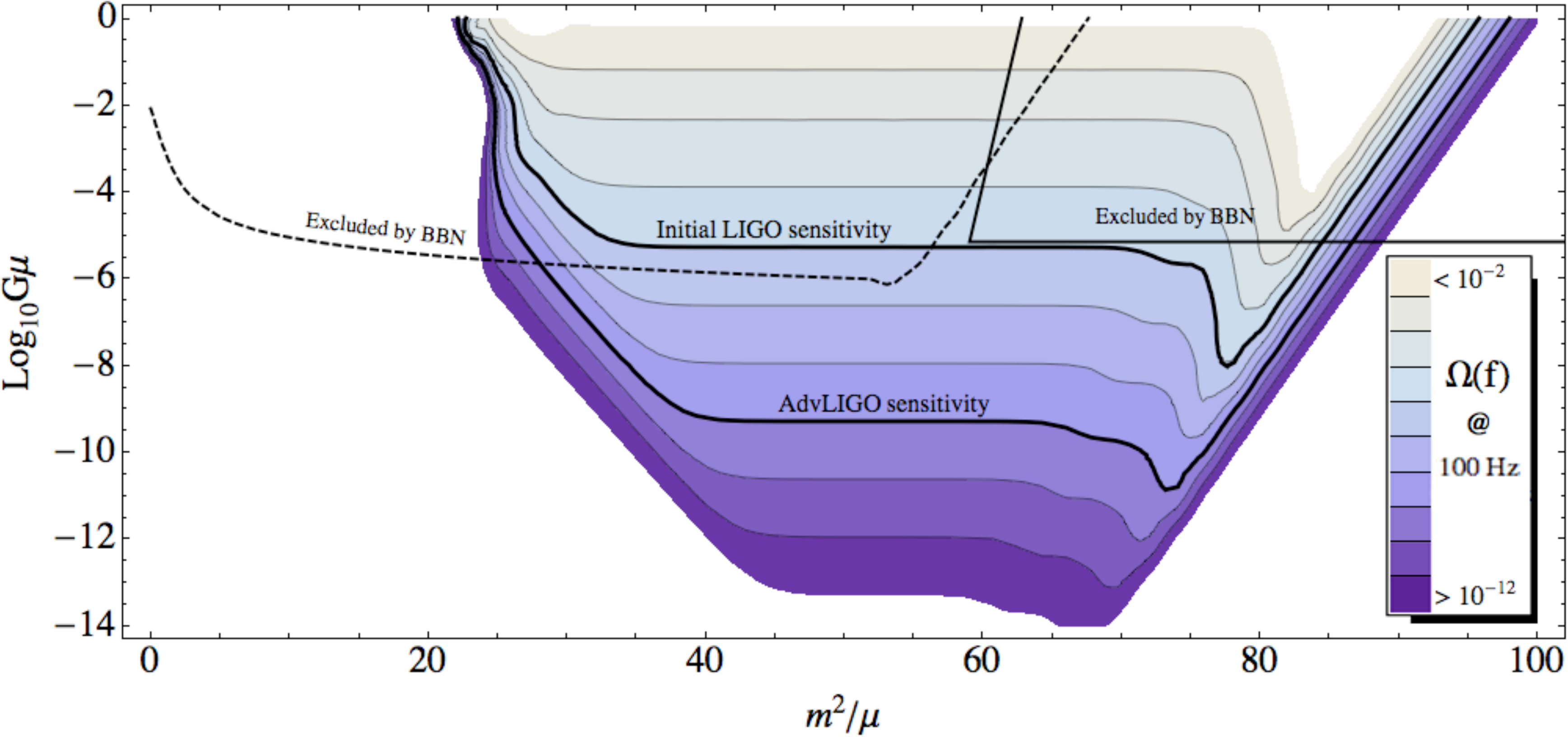}}
\caption{The upper graph shows the total density of gravitational waves. The value is nearly independent of $m$ for a large part of parameter space.  We have traced out the portion of the parameter space excluded by BBN from loops (solid curve) and beads (dashed curve).  Interestingly, BBN probes values down to the expected theoretical bound $m^2 \gtrsim \mu$.  Because the contours are linearly independent of the contours in the burst figures, it may be possible to determine both the bead mass and the string tension. On the lower graph we show the spectrum in the LIGO band.  The sensitivities of Initial and Advanced LIGO are also shown\cite{Abbott:2006zx}.}
\label{omegagw}
\end{figure}

\begin{figure}[h]
\centering{\includegraphics[width=6.4in]{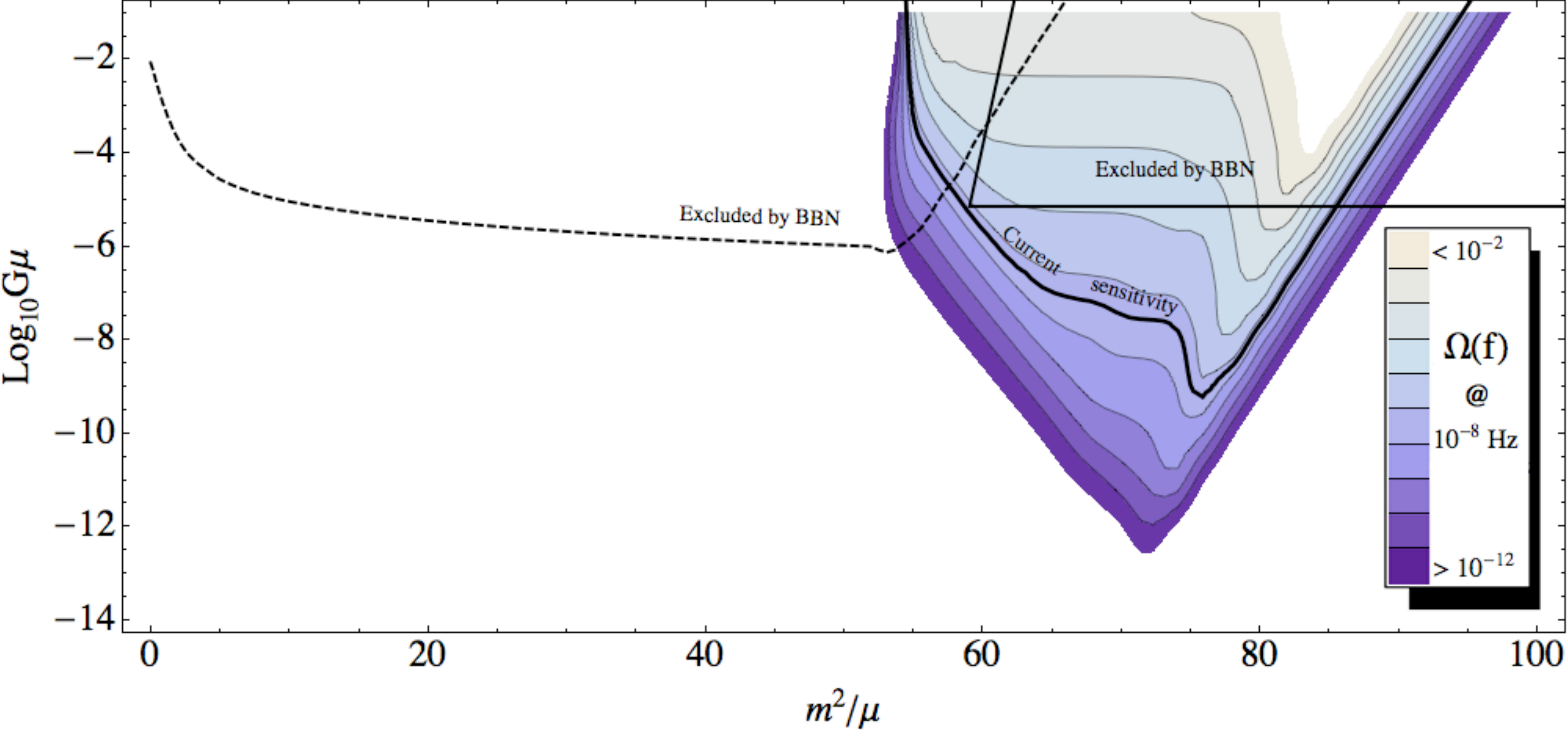}}
\caption{Spectrum in the pulsar band.  The current bound of $3.9 \times 10^{-8}$ is shown
\cite{Jenet:2006sv}.}
\label{stochasticpulsar}
\end{figure}

With some simple approximations, it is possible to determine the main properties of the stochastic background per logarithmic
frequency interval %(given by Eq.~(\ref{e:Omega(f)2}) )
\begin{equation}
\Omega_{\rm gw}(f) = \frac{4 \pi^2}{3H^2_0}f^3 
\int_{z_{**}}^{z_{max}} dz \int_{lmin}^{lmax} dl \, h^2(f,z,l)  \frac {dR}{dzdl}\; .
\end{equation}
The answer does not depend much on the upper bounds of the integrals and they can be taken to infinity.  The lower bounds are set by
\ba
z_{**} & \sim & \left(\frac{70}{H_0}\right)^{1/2} (\Gamma_2 \Gamma G\mu)^{1/4}\; ,\nonumber\\
l_{min} & \sim & \frac{1}{z f}\,\,.
\ea
For a decaying network of cosmic strings, all the energy density is transferred to gravitational waves.  If the decay is relatively sudden, the energy density in the gravitational waves is going to be dependent only on the energy in the network before decay, and is independent of the time at which the network decays.  We therefore expect the stochastic background to be only weakly dependent of $\kappa$, in the case when the network disintegrates during the radiation era.  In addition, due to the peculiar waveform associated with the ultra-relativistic beads, the spectrum is independent of frequency.
If the network always obeyed the scaling solution (with energy density $\rho_{cs} \propto G\mu/t^2$), then we would expect the stochastic background to be proportional to $G\mu$.  
Because short strings scale like matter, there is a deviation from scaling and there is a growth of string energy density for times between
the short string epoch ($t > t_*)$ and the demise of the network at $t_{**}$ (or $t_{eq}$, whichever comes first). 

The deviation from scaling between $t_*$ and $t_{**}$ provides an important factor of $\sqrt{t_{**}/t_*} = (\Gamma G\mu)^{-\tfrac{1}{4}}$ to the energy density.  %Because lighter strings radiate less, they will survive longer, and so further increase their relative energy density.  
Because of this, the stochastic background actually obeys
\be
\Omega_{\mbox{\tiny{gw}}} \propto (G\mu)^{3/4}\,\,,
\ee
which is an attractive feature of Fig.~(\ref{omegagw}) since it allows much lower tensions to produce an observable signal.   A similar effect is well known for cosmic string loops\cite{VilenkinShellard}.  We computed the total gravitational power and spectrum numerically, as shown in Fig.~(\ref{omegagw}) and (\ref{stochasticpulsar}).
The solid lines in all plots represents the region of parameter space for which radiation from string loops is in conflict with BBN.  In the simplest case, this restricts the string tension to satisfy $G\mu < 7\times 10^{-6}$ \cite{Caldwell:1991jj}.  Again, because no loops will exist for times after $t_*$, we have shown where this bound can no longer be reliably applied with the vertical kink in the solid curve.  Interestingly, when loops are eliminated before BBN, the tension remains constrained by the stochastic radiation from beads.  This is illustrated with the dashed line in all plots, defined by the BBN constraint
$\Omega_{\mbox{\tiny{gw}}}(z > z_{\mbox{\tiny{BBN}}}) < 1.5\times 10^{-5}$. Here, $\Omega_{\mbox{\tiny{gw}}}$ is defined as the quantity measured {\em today} from all sources more distant than $z_{\mbox{\tiny{BBN}}} = 5.5\times 10^9$. 

\begin{figure}[h]
\centering{\includegraphics[width=3.3in]{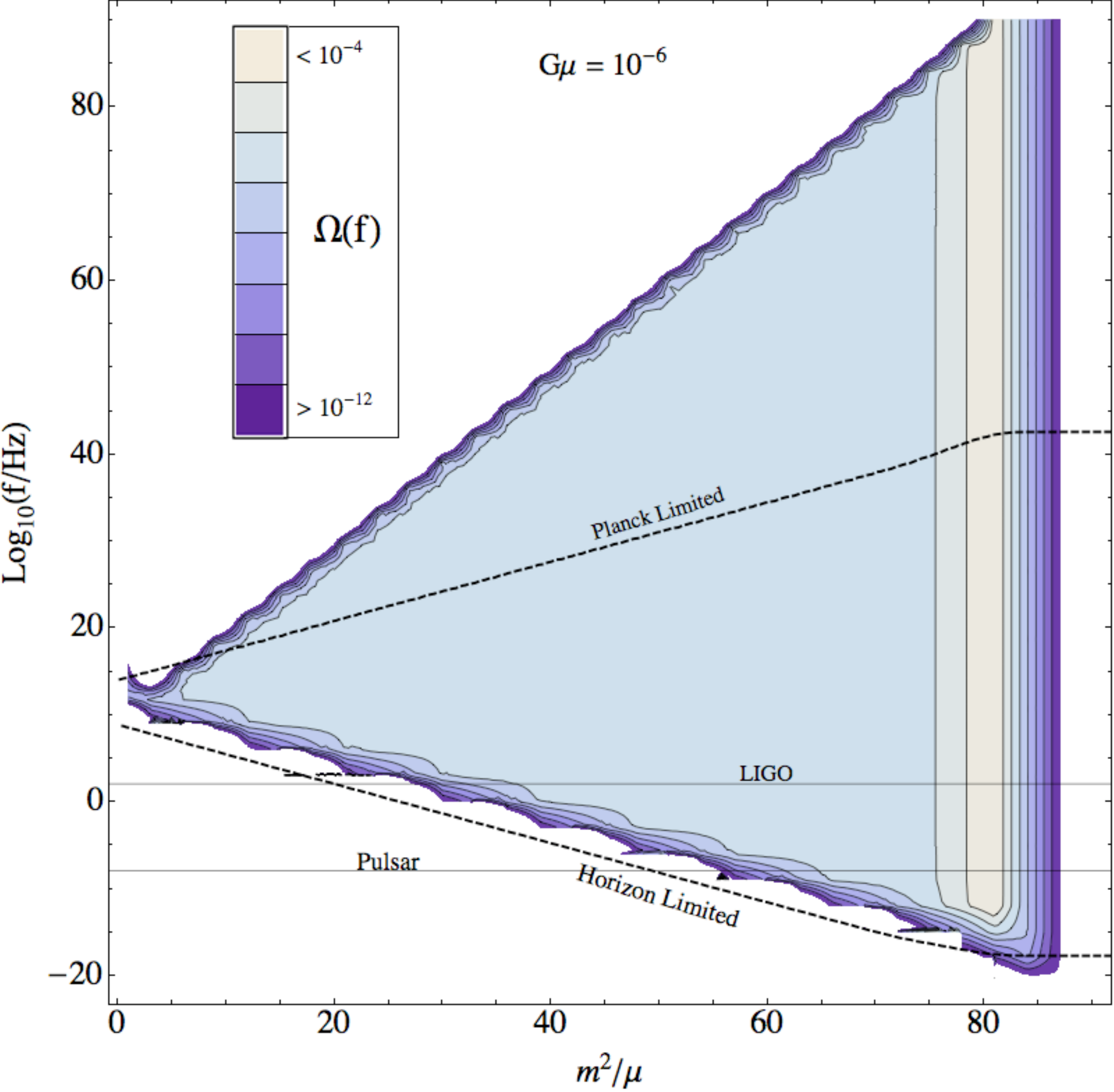}}
\centering{\includegraphics[width=3.3in]{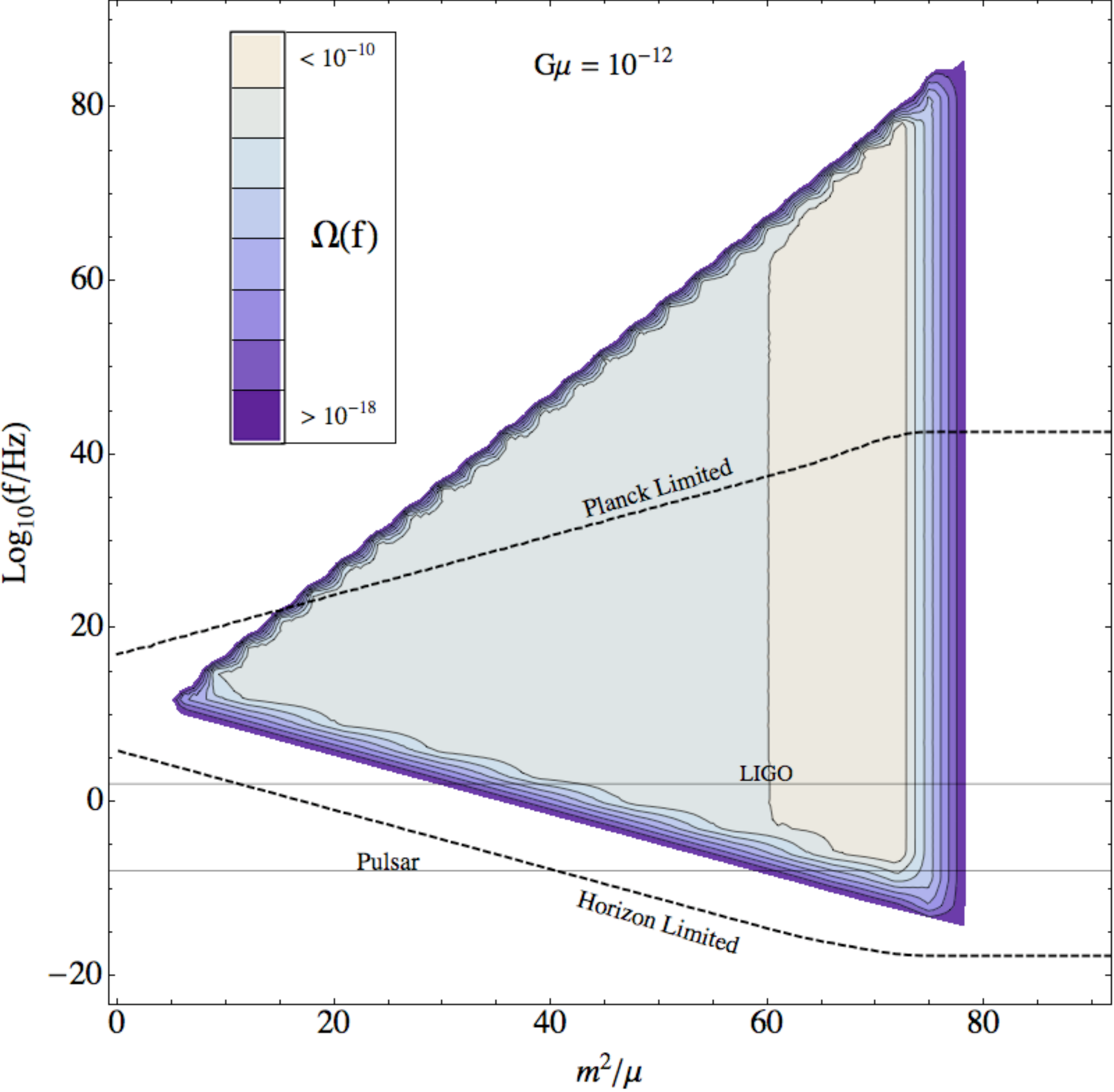}}
\caption{On the left we show the spectrum for $G\mu = 10^{-6}$. The spectrum is
scale invariant over wavelengths between the Hubble size and well beyond the Planck size.  Nearly half of the gravitational wave energy density was emitted at super-Planckian frequencies.  On the right we have the spectrum for $G\mu = 10^{-12}$. The IR and UV cutoffs are narrower than for high tension strings.  From this figure, it is clear that lower frequency experiments, {\em e.g.} pulsar timing, are ill-suited to probe a wide range of the bead mass, especially for low tension strings.}
\label{stochasticfreq}
\end{figure}

Not only are segments of all sizes radiating, but each burst spans a tremendous range of frequencies, from the length of the segment to well beyond the Planck scale.  Because of the scale invariance, the total power is simply expressed as
\be
\Omega_{\mbox{\tiny{gw}}} = \Omega_{\mbox{\tiny{gw}}}(f) \ln \frac{f_{max}}{f_{min}}
\ee
for any $f\in [f_{min},f_{max}]$, where the upper and lower bounds are related to the trans-Planckian and Hubble scales respectively, during the short string epoch, as is illustrated in Figs.~(\ref{stochasticfreq}). 
Of course our calculation breaks down past Planckian frequencies, and the fact that we have enough energy to excite these very high energy gravitons 
may indicate that massive, universally coupled fields (such as the dilaton in string theory) are important.   In any case, the ultra-high frequency background is totally decoupled from gravitational wave experiments. 

It can be understood why the lower tension strings will not produce the Hubble-limited long wavelengths found in the higher tension cases (compare the two cases in Fig.~\ref{stochasticfreq}).  Because the lower string tensions allow a longer period of scaling violation ({\em i.e.,} segments scaling like matter during the radiation era), the Hubble length grows much larger than the typical segment size during their radiation peak near $t_{**}$.  Similarly, the lighter strings do not bring the beads to the same speeds occurring for higher tension neworks, and the UV cutoff is lower as well.

\section{Conclusion}
The most model independent constraint on cosmic strings, whether meta-stable or stable, comes from BBN, which we find restricts the string
tension to satisfy ({\em c.f.} Fig.\ref{omegagw})
\be
G\mu \lesssim 10^{-5}
\ee
for {\em any} bead mass.  This can include strings which decay immediately after formation, and which are completely decoupled from further observations.  The remaining observables depend upon the degree to which cosmic strings are stable.
Theories of cosmic strings can only rarely claim the strings to be absolutely stable.  We find interesting phenomenology for meta-stable strings with bead mass within the range
\be\label{window}
1\,\,\lesssim \,\,\frac{m^2}{\mu} \,\,\lesssim \,\,100\,\,.
\ee
The degree to which this is a fine-tuning depends greatly on the mechanism responsible for string breakage.   For the bead mass range $40 \lesssim m^2/\mu \lesssim 80$, a stochastic background is detectable by  Advanced LIGO for tensions $G\mu \gtrsim 10^{-11}$. For the longer-lived strings with $70 \lesssim m^2/\mu \lesssim 80$, meta-stable strings provide a burst signal detectable by Advanced LIGO for string tensions as low as $G\mu \sim 10^{-12}$. The waveform of these bursts ($1/f$) is very distinctive, and a detection would confirm the ultra-relativistic nature of the source.  Already, initial LIGO rules out a relatively narrow range of $\kappa$ for tensions as low as $G\mu \sim 10^{-10}$.  A measurement of these bursts together with the stochastic background would enable one to determine both the string tension and the bead mass.  This can be seen from the linear-independence of the contours in Figs. (\ref{burst} \& \ref{omegagw}). 

An interesting feature of the predicted stochastic background is the degree to which it is scale invariant.  For higher tension strings, the spectrum extends between the Hubble scale (at the time of emission) to  the Planck scale, and possibly beyond.  In the upper range, one would expect universally coupled massive fields to be radiated with equal power as gravitational radiation \cite{Damour:1996pv,Babichev:2005qd}.  For lower tension strings, the dominant contribution to the gravitational power comes from shorter strings.  For this reason, low frequency experiments such as pulsar timing are ill-suited to constrain the parameter space of meta-stable cosmic strings.

Eventually, it may also be interesting to seek specific models within string theory where both the cosmic string tension and the value of $\kappa$ are predicted and presumably are related (models are known where $\kappa$ is much greater than one and is unrelated to the cosmic string tension).  We have only studied the case of a network of cosmic strings breaking due to beads, although it would be very interesting to study the phenomenological signals of the other possible network instability: a network of domain walls ending on cosmic strings.  We expect that gravitational waves will be produced with a similarly characteristic spectrum.  

\pagebreak
\bigskip

\centerline{\bf{Acknowledgements}}

\bigskip 
We would like to thank Jose Blanco-Pillado, Ken Olum, Amir Said, Alex
Vilenkin, and Alan Wiseman for useful discussions.  L.L. would like to thank the KITP
where this work was completed.
This research was supported in part by the National Science Foundation 
under Grant No. PHY05-51164, PHY-0758155, PHY-0505757.  The work of B.S. was supported in part by DOE Grant No. DE-FG02-91-ER-40672 and NSF Grant No. 0353314.

\begin{appendix}
\section{Decay Rate from the Path Integral}\label{sec:rate}
Let us calculate the string decay rate (following \cite{Coleman:1978ae}).  If we treat the string's ground state as an approximate energy eigenstate, the decay rate per unit length is equal to twice the imaginary part of the  energy density.  
The Euclidean action is given by
\baray
S_E &=& \mu\int_\Sigma dA \,\,+\,\, m\int_{\partial \Sigma} ds\\ 
&=& \mu XT - \mu\int_0^{2\pi}\hspace{-2mm}d\theta \frac{\rho^2(\theta)}{2} + m\int_0^{2\pi}\hspace{-2mm}d\theta \sqrt{\rho^2(\theta) + {\dot{\rho}}^2(\theta)}\,\,\,,
\earay
where the degree of freedom is the boundary, $\partial \Sigma \leftrightarrow  \rho(\theta)$.  We will
only consider the portion of $\Sigma$ that fits in a large rectangle of side lengths $X$ and $T$, and we assume the string is flat.
The path integral gives the vacuum energy density relation
\be
\left<0|\exp(-{\cal H} X T)|0\right> = N \int [{\cal D}\rho] \exp(-S_E[\rho])
\ee

We define $\mu$ by the perturbative result $\left<{\cal H}\right> = \mu$, and the leading
corrections to this can be calculated in the semiclassical dilute instanton gas approximation.
It can be readily shown that the above action has saddle points for
$\partial \Sigma$ equal to any number of non-overlaping circles
of radius $m/\mu$.  For a solution with $n$ such holes, the value of the action satisfies
\be
S_E[n\times S^1] - \mu X T =: nS_0  = n\pi m^2/\mu\,\,.
\ee
In the semi-classical approximation, we
expand the path integral around its saddle points.  The dilute
gas approximation is valid in the limit $X, T \to \infty$, so we ignore the overlap and write the path
integral equation as
\be
\left<e^{-{\cal H}XT}\right> = N \sum_{n=0}^\infty \int _{X^n,T^n}\frac{dx^ndt^n}{2^nn!} \frac{\exp\left(-nS_0 - \mu X T\right)}{2^n\sqrt{det'^n}}
\ee
We are summing over the number of instantons and their positions, dividing by the symmetry factor $n!$.  The quantity $det'$ is the determinant of the second functional variation of the action, representing fluctuations about a single instanton.  (We have set the perturbative contribution to the determinant to one by putting the physical parameters into the action.)  The prime indicates that we ignore the zero modes, since these correspond to the position, which has already been integrated over.  
The instanton is expanded in eigenfunctions of the second functional variation as
\ba
\rho(\theta) = m/\mu + \sum_{j\in {\mathbbm Z}}\frac{c_j}{\sqrt{\pi}} \left\{ \begin{array}{lr}\cos(j \theta)& j < 0\\1/\sqrt{2}& j = 0\\\sin(j \theta)& j > 0\end{array} \right.
\ea

The functional measure is 
\be
[{\cal D}\rho] \,\,\,=\,\,\, \prod_j \frac{dc_j}{\sqrt{2\pi}} \,\,\,=\,\,\, \frac{dx dt}{2}  \prod_j^{\qquad \prime} \hspace{-1mm} \frac{dc_j}{\sqrt{2\pi}}\,\,\,,
\ee
where the prime indicates we remove the zero modes ($j = \pm 1$).
The single negative mode $(j = 0)$ corresponding to dilatations ensures an imaginary contribution to the vacuum energy, {\em i.e.,} a non-zero decay rate.  Associated to this negative mode is an overall factor of $1/2$, since the path integral is divergent only for $c_0\to+\infty$, and not as $c_0\to-\infty$.  (This is the factor that Sagredo missed.)  
The result is
\be
\left<\exp(-{\cal H}XT)\right> = N\exp\left(-\frac{\tfrac{1}{2}X T e^{-S_0}}{2 \sqrt{det'}} - \mu X T\right)
\ee
with
\ba
det' &:=& det'\left[-\mu \partial_\theta^2 - \mu \right] \\&=& \hspace{-3mm} \prod_{j \in {\mathbbm {Z}}\, \,\,j \neq \pm1 } \hspace{-3mm}\mu (j^2 - 1) \,\,=\,\, -\pi^2/\mu^2\,\,.
\ea
We evaluated the determinant using the Riemann zeta function.  Hence
\be
\left<{\cal H}\right>  = \mu - i\frac{\mu}{4\pi}\exp\left(-\pi m^2/\mu\right)\,\,,
\ee
and so the bead pair production rate per unit length is
\be
\Gamma_2 = \frac{\mu}{2\pi}\exp\left(-\pi m^2/\mu\right)\,\,.
\ee

%%%%%%%%%%%%%%%%%%%%%%%%%%%%%%%%%%%%%%%%%%%%%%%%%%
%%%%%%%%%%%%%%%%%%%%%%%%%%%%%%%%%%%%%%%%%%%%%%%%%%
%%%%%%%%%%%%%%%%%%%%%%%%%%%%%%%%%%%%%%%%%%%%%%%%%%
%%%%%%%%%%%%%%%%%%%%%%%%%%%%%%%%%%%%%%%%%%%%%%%%%%
\section{Calculating the Number Density $n(l,t) dl$}\label{sec:distribution}
The number density of segments satisfies the Boltzmann equation
\begin{equation}
\label{eq:B:boltzmann}
\frac{\partial n(l,t)}{\partial t} = -\frac{\partial }{\partial
  l}\left(\dot l n(l,t)\right)- 3\frac{\dot a}{a} n(l,t) + g\,\,,
\ee

where we distinguish between three different source terms:

\be
g = g_{\tt loop} + g_{\tt ic}+ g_{\tt break}
\ee
with
\ba
\hspace{-3mm}g_{\tt break}dl &=&  \Gamma_2\left(2\int_l^\infty n(l',t)dl' \,\, - ln(l,t)\right)dl\\
g_{\tt ic}dl &=& \Gamma^{\tt ic}_{2}\left(\tfrac{1}{2}\int_{l/2}^\infty \hspace{-2mm}n(l',t)dl' \,\,-  l n(l,t)\right)dl\\
g_{\tt loop} &=& \mbox{(difficult)}
\ea

Thus there are three interactions:  Loop production is the process by which the network loses energy in the form of small loops which fragment and decay.  Intercommutation causes two segments to trade a quantity of string, keeping the total energy and number of segments fixed.  Breakage increases the number of segments but leaves the energy unaffected.

We motivated the breakage formula in the main text.  The intercommutation interaction $g_{ic}$ can be motivated as follows:  A string of length $l$ will intercommute at a rate equal to its length times the intercommutation rate per unit length, $\Gamma_2^{\tt ic}$.  (This rate is easily determined from the scaling solution, but will not be important for us.)  This explains the second term.  The creation of segments of length $l$ must involve a string longer than $l/2$, since we only need to consider two-string collisions.  This explains the integral, and the coefficient is over-constrained  by the fact that intercommutation should leave both the energy and number of segments constant, each of which demands the $\tfrac{1}{2}$.  Since intercommutation of segments preserves both the total number of segments and their average length, we will ignore its effect on $n(l,t)dl$.  (The essential feature of this interaction is it attempts to evolve $n(l,t)dl$ into a power-law distribution.  Intercommutation will extend the large $l$ tail of $n(l,t)dl$, but this is of little relevance for the short strings which actually radiate.)

Unfortunately, $g_{loop}$ is difficult to write down, since it depends strongly on how wiggly the strings are, which depends on how much string is present. 
Luckily, we can infer its value from the scaling solution.  
We expect the network will scale as long as most of the string is in superhorizon segments.
During scaling, two strong effects yield an attractor solution: stretching of superhorizon strings, and production of subhorizon loops.  Fewer strings cause fewer intercommutations, leading to straighter strings and lower loop production.  Stretching then causes the string energy density to grow toward the attractor value.  Too many strings will cause an increase in intercommutation events, making strings more wiggly, and more prone to loop production.  In the spirit of coarse graining, we will think of loop production simply as a  continuous shortening of string segments:
\be
g_{\tt loop} \to \dot{l}_{\tt loop}
\ee
A key feature is that a segment's length loss due to loop production is {\em proportional} to the segment length.  There are thus three important contributions to $\dot{l}$, namely Hubble stretching, gravitational radiation, and loop production.  
\be
\dot{l} = \dot{l}_H + \dot{l}_{\tt gw}+ \dot{l}_{\tt loop} 
\ee
Here we will {\em define} loop production as being consistent with the scaling solution, whereby the string energy density scales like $\rho_{cs} \sim \mu/t^2$.  Because $\dot{l}_{\tt loop} \propto l$, there is a unique way to do this.  

Gravitational radiation is only important for very short segments.  Scaling should cease when loop production becomes negligible.  This will happen if the strings become smooth, due to reduced intercommutation.  We expect such behavior during the short string epoch, when the network will be dominated by sub horizon sized segments with a density of a few per Hubble volume.  Since these segments will have slow peculiar velocities, intercommutation will be very rare.

Strings longer than twice the Hubble length are always dominated by potential energy, and so the
length (energy) is increased by expansion:
\be
\dot{l}_H = \frac{\dot{a}}{a} l (1 - 2 \left<v^2\right>) \qquad l \gg 2 H^{-1}
\ee
For long segments, $\left< v^2\right> \sim .4$, depending on how much intercommutation
is going on.  
If we set $g = 0$, the super-horizon string network would dominate the universe:
\be
\int_{}^{\infty} dl \mu l n(l,t)  \sim \frac{1}{a^{2 + 2 \left<v^2\right>}}
\ee
We know that the network will obey the scaling solution during the long string epoch.  Since
a scaling network obeys
\be\label{eq:rhocs}
\rho_{cs} = \int_0^\infty\mu l n(l,t)dl \quad \sim \quad \mu/t^2\,\,,
\ee
we can use this constraint to find the effective $\dot{l}$ simply by integrating Eqn.(\ref{eq:boltzmann}).
The result is
\ba
\dot{l}_{\tt} &=& \dot{l}_H + \dot{l}_{\tt loop} + \dot{l}_{\tt gw}\\
&=& 3 H l - \frac{2 l}{t}\quad\quad{\mbox{(scaling solution)}}
\ea
As discussed in the text, the scaling solution will apply for $t < t_*$, after which
\be
\dot{l}\,\, \approx \,\, -\Gamma G\mu\,\,.
\ee
With an explicit expression for $\dot{l}$, we can then solve the Boltzmann equation exactly, obtaining
\ba
n(l,t) &=&  \left\{ \begin{array}{ll} (2 \Gamma_2)^2e^{-2 \Gamma_2 l t} &  \;\;\; t \leq t_*\leq t_{eq}\\[2mm] 
\frac{(t + t_*)^2\Gamma_2^2}{\sqrt{t^3 t_*}} e^{-\Gamma_2 l(t + t_*)}e^{-(t - t_*)(t + 3 t_*)/2 t_{**}^2} &  \;\;\; t_* \leq t \leq t_{eq}\\[2mm]
\sqrt{\frac{t_{eq}}{t_*}}\frac{(t + t_*)^2\Gamma_2^2}{t^2} e^{-\Gamma_2 l(t + t_*)}e^{-(t - t_*)(t + 3 t_*)/2 t_{**}^2}
 & \;\;\;  t_* \leq t_{eq} \leq t\end{array} \right.
\ea
for $t_* \leq t_{eq}$, and 
\ba
n(l,t) &=&  \left\{ \begin{array}{ll} 
(2 \Gamma_2)^2e^{-2 \Gamma_2 l t} & \;\;\; t \leq t_{eq}\leq t_*\\[2mm]
 \Gamma_2^2 \frac{\left(t + t_{eq}\right)^2}{t^2}e^{-\Gamma_2l\left(t + t_{eq}\right)} &\;\;\;  t_{eq} \leq t \leq t_*\\[2mm]
\frac{\left(t + t_{eq}\right)^2}{t^2}\Gamma_2^2 e^{-\Gamma_2 l (t + t_{eq})} e^{-\left(t-t_*\right)\left(t + 2 t_{eq}\right)/2t^2_{**}} & \;\;\; t_{eq}\leq t_*\leq t\end{array} \right.
\ea
for $t_{eq} \leq t_*$.

\section{Gauge Invariant Degrees of Freedom}
\label{gaugeinv}

In order to obtain the gauge invariant part of the gravitational wave
we only need consider the spatial components
\begin{eqnarray}
\label{hij} 
\epsilon_{ij} ({\bf x},\omega)&=&
\frac{2G}{r}T^{03}(k)
\left( 
\begin{array}{ccc}
\cos\theta-1/\cos\theta & 0 & 0 \\
0 & \cos\theta-1/\cos\theta & 0 \\
0 & 0 & \cos\theta+1/\cos\theta\\
\end{array}
\right)
\end{eqnarray}

The transverse-traceless part of the polarization tensor can be obtained using the 
projection operator \cite{MTW}
\begin{equation}
P_{ij}=\delta_{ij}-{\hat \Omega}_i{\hat \Omega}_j,
\label{proj1} 
\end{equation}
as follows
\begin{equation}
\epsilon^{TT}=P \epsilon P-\tfrac{1}{2}Tr(P\epsilon).
\label{htt1} 
\end{equation}

If the wave were to lie entirely on the $z$-axis the projector would be
\begin{equation}
P_{ij}=\left( 
\begin{array}{ccc}
1& 0 & 0 \\
0 & 1 & 0 \\
0 & 0 & 0\\
\end{array}
\right).
\label{proj2} 
\end{equation}
So that some arbitrary polarization tensor
\begin{equation}
\epsilon_{ij}=\left( 
\begin{array}{ccc}
\epsilon_{11}& \epsilon_{12} & \epsilon_{13} \\
\epsilon_{21}& \epsilon_{22} & \epsilon_{23} \\
\epsilon_{31}& \epsilon_{32} & \epsilon_{33} \\
\end{array}
\right).
\label{proj3} 
\end{equation}
becomes
\begin{equation}
\epsilon^{TT}_{ij}=\left( 
\begin{array}{ccc}
(\epsilon_{11}-\epsilon_{22})/2& \epsilon_{12} & 0 \\
\epsilon_{21}& (\epsilon_{22}-\epsilon_{11})/2 & 0 \\
0 & 0 & 0 \\
\end{array}
\right).
\label{proj4} 
\end{equation}
The so called $+$- and $\times$-polarisations are defined by
\begin{equation}
\label{ep}
\epsilon_{+} \equiv \tfrac{1}{2} (\epsilon_{11}-\epsilon_{22})
\end{equation}
and
\begin{equation}
\label{ec}
\epsilon_{\times} \equiv  \epsilon_{12}= \epsilon _{21}.
\end{equation}

For a wave not traveling in the $z$-direction we can perform a rotation so that the 
wave vector does lie on the $z$-axis and then apply the arguments outlined above. 
This can easily be done noting that
\begin{equation}
\label{la1}
R_x(\theta)R_z(\frac{\pi}{2}-\phi){\hat \Omega}={\hat z}.
\end{equation}
The rotated metric perturbation polarization tensor is therefore
\begin{equation}
\epsilon'=R_x(\theta)R_z(\frac{\pi}{2}-\phi) \epsilon R^T _z(\frac{\pi}{2}-\phi) R^T _x(\theta),
\label{proj5} 
\end{equation}
with the rotation matrices defined by
\begin{equation}
R_{z}(a)=\left( \begin{array}{ccc}
\cos a & -\sin a & 0 \\
\sin a &  \cos a & 0 \\
0 & 0 & 1
\end{array} \right),
\end{equation}
and
\begin{equation}
R_{x}(a)=\left( \begin{array}{ccc}
1 & 0 & 0 \\
0 & \cos a & -\sin a \\
0 & \sin a &  \cos a 
\end{array} \right)
\end{equation}
as usual.

If we perform the rotation Eq.~(\ref{proj5}) on Eq.~(\ref{hij}) we obtain 
\begin{eqnarray}
\label{hijp} 
\epsilon_{ij} ({\bf x},\omega)&=&
\frac{2Gm}{r} T^{03}(k)
\left (
\begin {array}{ccc} 
{\frac {\cos^2\theta-1}{\cos\theta}}&0&0\\
0& -\frac {\cos^2\theta-1}{\cos\theta}&-2\sin\theta\\
0&-2\sin\theta&
{\frac {3 \cos^2\theta-1}{\cos\theta}}\end {array}\right ).
\end{eqnarray}
The transverse traceless part is therefore
\begin{eqnarray}
\label{hijpTT} 
\epsilon ^{TT}_{ij} ({\bf x},\omega)&=&
\frac{2Gm}{r}{\frac {\cos^2\theta-1}{\cos\theta}}T^{03}(k)
\left (
\begin {array}{ccc} 
1&0&0\\
0&-1&0\\
0&0&0
\end {array}\right )
\end{eqnarray}
from which the $+$- and $\times$-polarisations can easily be read off,
\begin{equation}
\label{hplus2}
\epsilon_+=\frac{2Gm}{r} \frac {\cos^2\theta-1}{\cos\theta} T^{03}(k)
\end{equation}
and
\begin{equation}
\label{hcross2}
\epsilon_\times=0.
\end{equation}

%%%%%%%%%%%%%%%%%%%%%%%%%%%%%%%%%%%%%%%%%%%%%%%
%%%%%%%%%%%%%%%%%%%%%%%%%%%%%%%%%%%%%%%%%%%%%%%%
%%%%%%%%%%%%%%%%%%%%%%%%%%%%%%%%%%%%%%%%%%%%%%%%
%%%%%%%%%%%%%%%%%%%%%%%%%%%%%%%%%%%%%%%%%%%%%%%%

%%%%%%%%%%%%%%%%%%%%%%%%%%%%%%%%%%%%%%%%%%%%%
%%%%%%%%%%%%%%%%%%%%%%%%%%%%%%%%%%%%%%%%%%%%%
%%%%%%%%%%%%%%%%%%%%%%%%%%%%%%%%%%%%%%%%%%%%%
%%%%%%%%%%%%%%%%%%%%%%%%%%%%%%%%%%%%%%%%%%%%%

\section{Gravitational Waveform at High Frequency}
\label{continuous}
Here we compute the gravitational waveform produced by a
discontinuity in the bead acceleration which can occur when
beads cross on a straight string or when a bead encounters a kink
on a string.  We start from the results of Martin and
Vilenkin~\cite{MV} and calculate the gravitational waveform 
for a fly-by of a bead - antibead pair using a
Taylor expansion around the time of the crossing.
We start with the stress-energy tensor given by Eq.~(\ref{Tmunu01}) but
rather than using Eq.~(\ref{x1}), we can Taylor expand $x_1(t)$ around $t=0$
writing $x_1(t)\approx v_0 t$.  The Fourier transform of Eq.~(\ref{Tmunu01}) 
then reads
\begin{equation}
\label{TmunuFT3cont} 
T^{03}(k)=\frac{m}{2\pi} \int_{-\infty}^{\infty} dt 
(\gamma_0 v_0 -a|t|) \left[ e^{i \omega_- t} 
-  e^{i \omega_+ t} \right],
\end{equation}
where $\omega_{\pm}=\omega(1\pm v_0\cos\theta)$.  This integral can be
done trivially.  The integral,
\begin{equation}
\label{FTintegral} 
\frac{m}{2\pi} \int_{-\infty}^{\infty} dt 
(\gamma_0 v_0 -a|t|) e^{i \omega t}=\frac{am}{\sqrt{2 \pi} \pi \omega^2},
\end{equation}
where we've thrown out a static field piece proportional to
$\delta(\omega)$.  This means,
\begin{equation}
\label{TmunuFT4cont} 
T^{03}(k)=\frac{am}{\sqrt{2 \pi} \pi}\left[\frac{1}{\omega_-^2}-\frac{1}{\omega_+^2}\right].
\end{equation}
Similarly as in the main text, the gravitational 
wave is linearly polarized and the plus polarization can be written as
\begin{equation}
\label{polariz3cont2}
\epsilon_{+}  ({\bf x},\omega)=\frac{2G}{r}\frac{am}{\sqrt{2 \pi}
  \pi}\left[\frac{1}{\omega_-^2}-\frac{1}{\omega_+^2}\right]
\frac{\cos^2\theta-1}{\cos\theta}
\end{equation}
If the beads are relativistic the radiation is beamed in the
direction of the bead velocity.  If the direction of observation is
along that direction then one of the $1/\omega_-^2$ or $1/\omega_+^2$
is large and the other is $O(1)$, {\em i.e.,} we are only observing the
radiation from one of the two beads.  If we choose the direction so
that $\cos\theta\approx 1$, since $a=\mu/m$ we can write the waveform
as
\begin{equation}
\label{polariz4cont}
\epsilon_{+}  ({\bf x},\omega)=\frac{2G\mu}{r\sqrt{2 \pi}
  \pi\omega^2}\frac{1}{(1-v_0\cos\theta)^2}\frac{\cos^2\theta-1}{\cos\theta}
\end{equation}
In the region where most of the energy is radiated,
i.e.  Angles $\theta \sim 1/\gamma_0$ with $\gamma_0 \gg 1$, we can
use the small angle approximation and set $v_0\approx 1$ to obtain,
\begin{equation}
\label{polariz5cont}
\epsilon_{+}  ({\bf x},\omega) \approx 
\frac{2G\mu}{r\sqrt{2 \pi}\pi}\left(\frac{\gamma_0}{\omega}\right)^2.
\end{equation}
Note that although we have done the computation for the fly-by of a
bead - antibead pair the waveform is valid up to factors of
${\cal O}(1)$ for a bead encounter with a kink on the string which
also produces a discontinuity in the acceleration.  In the case of a
fly-by, if the beads miss each other with impact parameter $b$, or
for a bead encounter with a kink smoothed on such a length scale,
we expect the waveform to be suppressed at frequencies above $b^{-1}$.
For a bead - antibead pair connected by a string of length $l$
we expect the impact parameter $b$ to be of ${\cal O}(l)$, so unless
the strings are very straight most of the signal will be at
frequencies near $l^{-1}$.  On the other hand kinks smooth
gravitationally very slowly, remaining sharp for the lifetime of the
bead - antibead system.  Therefore beads connected by a
kinky string will encounter sharp kinks and radiate every oscillation
for the lifetime of the system.  
Neglecting gravitational wave back-reaction, the kink-bead collision can be
solved exactly.  We can write the strain at frequency $f$ for a burst produced at a distance $r$, 
\begin{equation}
\label{strain3cont}
h(f,l,z) \approx 
\frac{G\mu H_0}{(1+z)\varphi_r(z)}\left(\frac{\mu l}{m f}\right)^2
\end{equation}
The amplitude
is suppressed by an extra factor of $(1+z)$ due to the redshifting of
the frequency $f$ (see the discussion in section IV of~\cite{DV1}). 

\end{appendix}


\begin{thebibliography}{unsrt}

\bibitem{Langacker}
  P.~Langacker and S.~Y.~Pi,
  ``Magnetic Monopoles In Grand Unified Theories,''
  Phys.\ Rev.\ Lett.\  {\bf 45}, 1 (1980).
  %%CITATION = PRLTA,45,1;%%


\bibitem{bands}
  P.~Bhattacharjee and G.~Sigl,
  ``Origin and propagation of extremely high energy cosmic rays,''
  Phys.\ Rept.\  {\bf 327}, 109 (2000)
  [arXiv:astro-ph/9811011].
  %%CITATION = PRPLC,327,109;%%

\bibitem{berez}
  V.~Berezinsky, B.~Hnatyk and A.~Vilenkin,
  ``Gamma ray bursts from superconducting cosmic strings,''
  Phys.\ Rev.\  D {\bf 64}, 043004 (2001)
  [arXiv:astro-ph/0102366].
  %%CITATION = PHRVA,D64,043004;%%

\bibitem{Vachaspati:2008su}
  T.~Vachaspati,
   ``Cosmic Sparks from Superconducting Strings,''
  arXiv:0802.0711 [astro-ph].
  %%CITATION = ARXIV:0802.0711;%%

\bibitem{Battefeld:2007qn}
  D.~Battefeld, T.~Battefeld, D.~H.~Wesley and M.~Wyman,
  ``Magnetogenesis from Cosmic String Loops,''
  JCAP {\bf 0802}, 001 (2008)
  [arXiv:0708.2901 [astro-ph]].

%\cite{Mack:2007ae}
\bibitem{Mack:2007ae}
  K.~J.~Mack, D.~H.~Wesley and L.~J.~King,
  ``Observing cosmic string loops with gravitational lensing surveys,''
  Phys.\ Rev.\  D {\bf 76}, 123515 (2007)
  [arXiv:astro-ph/0702648].
  %%CITATION = PHRVA,D76,123515;%%

%\cite{Gasparini:2007jj}
\bibitem{Gasparini:2007jj}
  M.~A.~Gasparini, P.~Marshall, T.~Treu, E.~Morganson and F.~Dubath,
  ``Direct Observation of Cosmic Strings via their Strong Gravitational Lensing
  Effect: I. Predictions for High Resolution Imaging Surveys,''
  arXiv:0710.5544 [astro-ph].
  %%CITATION = ARXIV:0710.5544;%%

\bibitem{Christiansen:2008vi}
  J.~L.~Christiansen, E.~Albin, K.~A.~James, J.~Goldman, D.~Maruyama and 
G.~F.~Smoot,
``Search for Cosmic Strings in the GOODS Survey,''
  Phys.\ Rev.\  D {\bf 77}, 123509 (2008)
  [arXiv:0803.0027 [astro-ph]].


\bibitem{Dyda:2007su}
  S.~Dyda and R.~H.~Brandenberger,
  ``Cosmic Strings and Weak Gravitational Lensing,''
  arXiv:0710.1903 [astro-ph].


\bibitem{Chernoff:2007pd}
  D.~F.~Chernoff and S.~H.~H.~Tye,
  ``Cosmic String Detection via Microlensing of Stars,''
  arXiv:0709.1139 [astro-ph].

%\cite{Kuijken:2007ma}
\bibitem{Kuijken:2007ma}
  K.~Kuijken, X.~Siemens and T.~Vachaspati,
  ``Microlensing by Cosmic Strings,''
  arXiv:0707.2971 [astro-ph].
  %%CITATION = ARXIV:0707.2971;%%

   %\cite{Baumann:2008aj}
\bibitem{Baumann:2008aj}
  D.~Baumann {\it et al.},
  ``CMBPol Mission Concept Study: A Mission to Map our Origins,''
  arXiv:0811.3911 [astro-ph].
  %%CITATION = ARXIV:0811.3911;%%


%\cite{Fraisse:2007nu}
\bibitem{Fraisse:2007nu}
  A.~A.~Fraisse, C.~Ringeval, D.~N.~Spergel and F.~R.~Bouchet,
  ``Small-Angle CMB Temperature Anisotropies Induced by Cosmic Strings,''
  Phys.\ Rev.\  D {\bf 78}, 043535 (2008)
  [arXiv:0708.1162 [astro-ph]].
  %%CITATION = PHRVA,D78,043535;%%

\bibitem{Pogosian:2008am}
  L.~Pogosian, S.~H.~Tye, I.~Wasserman and M.~Wyman,
  ``Cosmic Strings as the Source of Small-Scale Microwave Background
  Anisotropy,''
   JCAP {\bf 0902}, 013 (2009)
  [arXiv:0804.0810 [astro-ph]].
  %%CITATION = JCAPA,0902,013;%%

 
\bibitem{Khatri:2008zw}
  R.~Khatri and B.~D.~Wandelt,
  ``Cosmic (super)string constraints from 21 cm radiation,''
  Phys.\ Rev.\ Lett.\  {\bf 100}, 091302 (2008)
  [arXiv:0801.4406 [astro-ph]].


%\cite{Damour:2001bk}
\bibitem{DV1}
  T.~Damour and A.~Vilenkin,
  ``Gravitational wave bursts from cusps and kinks on cosmic strings,''
  Phys.\ Rev.\  D {\bf 64}, 064008 (2001)
  [arXiv:gr-qc/0104026].
  %%CITATION = PHRVA,D64,064008;%%

 
 \bibitem{SCMMCR} 
  X.~Siemens, J.~Creighton, I.~Maor, S.~Ray Majumder, K.~Cannon and J.~Read,
  ``Gravitational wave bursts from cosmic (super)strings: Quantitative
  analysis and constraints,''
  Phys.\ Rev.\  D {\bf 73}, 105001 (2006)
  [arXiv:gr-qc/0603115].
  %%CITATION = PHRVA,D73,105001;%%


 
 %\cite{Jenet:2006sv}
\bibitem{Jenet:2006sv}
  F.~A.~Jenet {\it et al.},
  ``Upper bounds on the low-frequency stochastic gravitational wave  background from pulsar timing observations: Current limits and future  prospects,''
  Astrophys.\ J.\  {\bf 653}, 1571 (2006)
  [arXiv:astro-ph/0609013].
  %%CITATION = ASJOA,653,1571;%%
 \bibitem{SMC}  X.~Siemens, V.~Mandic and J.~Creighton,
  ``Gravitational wave stochastic background from cosmic (super)strings,''
  Phys.\ Rev.\ Lett.\  {\bf 98}, 111101 (2007)
  [arXiv:astro-ph/0610920].
  %%CITATION = PRLTA,98,111101;%%

 

  %\cite{Jones:2002cv}
\bibitem{Tye}
  N.~T.~Jones, H.~Stoica and S.~H.~H.~Tye,
  ``Brane interaction as the origin of inflation,''
  JHEP {\bf 0207}, 051 (2002)
  [arXiv:hep-th/0203163].
 %%CITATION = JHEPA,0207,051;%%
 

 
 %\cite{Sarangi:2002yt}
%\bibitem{Sarangi:2002yt}
  S.~Sarangi and S.~H.~H.~Tye,
  ``Cosmic string production towards the end of brane inflation,''
  Phys.\ Lett.\  B {\bf 536}, 185 (2002)
  [arXiv:hep-th/0204074].
  %%CITATION = PHLTA,B536,185;%%
    
      %\cite{Copeland:2003bj}
\bibitem{CMP}
  E.~J.~Copeland, R.~C.~Myers and J.~Polchinski,
  ``Cosmic F- and D-strings,''
  JHEP {\bf 0406}, 013 (2004)
  [arXiv:hep-th/0312067].
  %%CITATION = JHEPA,0406,013;%%

  %\cite{Firouzjahi:2006vp}
\bibitem{Firouz}
  H.~Firouzjahi, L.~Leblond and S.~H.~Henry Tye,
  ``The (p,q) string tension in a warped deformed conifold,''
  JHEP {\bf 0605}, 047 (2006)
  [arXiv:hep-th/0603161].
  %%CITATION = JHEPA,0605,047;%%

%\cite{Jackson:2004zg}
\bibitem{Jack}
  M.~G.~Jackson, N.~T.~Jones and J.~Polchinski,
  ``Collisions of cosmic F- and D-strings,''
  JHEP {\bf 0510}, 013 (2005)
  [arXiv:hep-th/0405229].
  %%CITATION = JHEPA,0510,013;%%

\bibitem{Jackson:2009fk}
  M.~G.~Jackson and X.~Siemens,
  ``Gravitational Wave Bursts from Cosmic Superstring Reconnections,''
  arXiv:0901.0867 [hep-th].
  %%CITATION = ARXIV:0901.0867;%%


%\cite{Shlaer:2005ry}
\bibitem{Shlaer:2005ry}
  B.~Shlaer and M.~Wyman,
  ``Cosmic superstring gravitational lensing phenomena: Predictions for networks of (p,q) strings,''
  Phys.\ Rev.\  D {\bf 72}, 123504 (2005)
  [arXiv:hep-th/0509177].
  %%CITATION = PHRVA,D72,123504;%%  

  %\cite{Witten:1985fp}
\bibitem{Witten}
  E.~Witten,
  ``Cosmic Superstrings,''
  Phys.\ Lett.\  B {\bf 153}, 243 (1985).
  %%CITATION = PHLTA,B153,243;%%

%\cite{Leblond:2004uc}
\bibitem{LebTye}
  L.~Leblond and S.~H.~H.~Tye,
  ``Stability of D1-strings inside a D3-brane,''
  JHEP {\bf 0403}, 055 (2004)
  [arXiv:hep-th/0402072].
  %%CITATION = JHEPA,0403,055;%%
    
  %\cite{Callan:1997kz}
\bibitem{Callan}
  C.~G.~Callan and J.~M.~Maldacena,
  ``Brane dynamics from the Born-Infeld action,''
  Nucl.\ Phys.\  B {\bf 513}, 198 (1998)
  [arXiv:hep-th/9708147].
  %%CITATION = NUPHA,B513,198;%%
  
%\cite{Verlinde:2006bc}
\bibitem{Verlinde}
  H.~Verlinde,
  ``On metastable branes and a new type of magnetic monopole,''
  arXiv:hep-th/0611069.
  %%CITATION = HEP-TH/0611069;%%
  
  %\cite{Gubser:1998fp}
\bibitem{Gubser}
  S.~S.~Gubser and I.~R.~Klebanov,
  ``Baryons and domain walls in an N = 1 superconformal gauge theory,''
  Phys.\ Rev.\  D {\bf 58}, 125025 (1998)
  [arXiv:hep-th/9808075].
  %%CITATION = PHRVA,D58,125025;%%

%\cite{Leblond:2007tf}
\bibitem{LebWym}
  L.~Leblond and M.~Wyman,
  ``Cosmic Necklaces from String Theory,''
  Phys.\ Rev.\  D {\bf 75}, 123522 (2007)
  [arXiv:astro-ph/0701427].
  %%CITATION = PHRVA,D75,123522;%%
  
     %\cite{BlancoPillado:2007zr}
\bibitem{Blanco}
  J.~J.~Blanco-Pillado and K.~D.~Olum,
  ``Monopole annihilation in cosmic necklaces,''
  [arXiv:0707.3460].
  %%CITATION = ARXIV:0707.3460;%%  
  
  
      %\cite{BlancoPillado:1999cy}
\bibitem{BlancoPillado:1999cy}
  J.~J.~Blanco-Pillado and K.~D.~Olum,
  ``Monopole antimonopole bound states as a source of ultra-high-energy  cosmic rays,''
  Phys.\ Rev.\  D {\bf 60}, 083001 (1999)
  [arXiv:astro-ph/9904315].
  %%CITATION = PHRVA,D60,083001;%% 
  
 %\cite{Berezinsky:1997td}
\bibitem{Berezinsky:1997td}
  V.~Berezinsky and A.~Vilenkin,
  ``Cosmic necklaces and ultrahigh energy cosmic rays,''
  Phys.\ Rev.\ Lett.\  {\bf 79}, 5202 (1997)
  [arXiv:astro-ph/9704257].
  %%CITATION = PRLTA,79,5202;%%
  
%\cite{Preskill:1992ck}
\bibitem{PreskillVilenkin}
  J.~Preskill and A.~Vilenkin,
  ``Decay of metastable topological defects,''
  Phys.\ Rev.\  D {\bf 47}, 2324 (1993)
  [arXiv:hep-ph/9209210].
  %%CITATION = PHRVA,D47,2324;%%

%\cite{Jeannerot:2003qv}
\bibitem{Jeannerot:2003qv}
  R.~Jeannerot, J.~Rocher and M.~Sakellariadou,
  ``How generic is cosmic string formation in SUSY GUTs,''
  Phys.\ Rev.\  D {\bf 68}, 103514 (2003)
  [arXiv:hep-ph/0308134].
  %%CITATION = PHRVA,D68,103514;%%

  %\cite{Martin:1996ea}
\bibitem{Martin:1996ea}
  X.~Martin and A.~Vilenkin,
  ``Gravitational wave background from hybrid topological defects,''
  Phys.\ Rev.\ Lett.\  {\bf 77}, 2879 (1996)
  [arXiv:astro-ph/9606022].
  %%CITATION = PRLTA,77,2879;%%
  
  %\cite{Vilenkin:1982hm}
\bibitem{Vilenkin:1982hm}
  A.~Vilenkin,
  ``Cosmological Evolution Of Monopoles Connected By Strings,''
  Nucl.\ Phys.\  B {\bf 196}, 240 (1982).
  %%CITATION = NUPHA,B196,240;%%

 
%\cite{Monin:2008mp}
\bibitem{Monin:2008mp}
  A.~Monin and M.~B.~Voloshin,
  ``The spontaneous breaking of a metastable string,''
  Phys.\ Rev.\  D {\bf 78}, 065048 (2008)
  [arXiv:0808.1693 [hep-th]].
  %%CITATION = PHRVA,D78,065048;%%
  %\cite{Monin:2009ch}
%\bibitem{Monin:2009ch}
  A.~Monin and M.~B.~Voloshin,
  ``Destruction of a metastable string by particle collisions,''
  arXiv:0902.0407 [hep-th].
  %%CITATION = ARXIV:0902.0407;%%


%\cite{Martin:1996cp}
\bibitem{MV}
  X.~Martin and A.~Vilenkin,
  ``Gravitational radiation from monopoles connected by strings,''
  Phys.\ Rev.\  D {\bf 55}, 6054 (1997)
  [arXiv:gr-qc/9612008].
  %%CITATION = PHRVA,D55,6054;%%
  
  %\cite{Damour:1996pv}
\bibitem{Damour:1996pv}
  T.~Damour and A.~Vilenkin,
  ``Cosmic strings and the string dilaton,''
  Phys.\ Rev.\ Lett.\  {\bf 78}, 2288 (1997)
  [arXiv:gr-qc/9610005].
  %%CITATION = PRLTA,78,2288;%%
  
  %\cite{Babichev:2005qd}
\bibitem{Babichev:2005qd}
  E.~Babichev and M.~Kachelriess,
  ``Constraining cosmic superstrings with dilaton emission,''
  Phys.\ Lett.\  B {\bf 614}, 1 (2005)
  [arXiv:hep-th/0502135].
  %%CITATION = PHLTA,B614,1;%%



  \bibitem{VilenkinShellard}
Vilenkin A., Shellard E.P.S, 1994, {\em Cosmic strings and other
topological defects}.  Cambridge Univ.Press., Cambridge.


  %\cite{Polchinski:2005bg}
\bibitem{Open}
  J.~Polchinski,
  ``Open heterotic strings,''
  JHEP {\bf 0609}, 082 (2006)
  [arXiv:hep-th/0510033].
  %%CITATION = JHEPA,0609,082;%%


%\cite{Kachru:2003sx}
\bibitem{KKLMMT}
  S.~Kachru, R.~Kallosh, A.~Linde, J.~M.~Maldacena, L.~P.~McAllister and S.~P.~Trivedi,
  ``Towards inflation in string theory,''
  JCAP {\bf 0310}, 013 (2003)
  [arXiv:hep-th/0308055].
  %%CITATION = JCAPA,0310,013;%%

 
\bibitem{D3vacuua}
%%\cite{DeWolfe:2007hd}
%\bibitem{DeWolfe:2007hd}
  O.~DeWolfe, L.~McAllister, G.~Shiu and B.~Underwood,
  ``D3-brane Vacua in Stabilized Compactifications,''
  JHEP {\bf 0709}, 121 (2007)
  [arXiv:hep-th/0703088].
%  %%CITATION = JHEPA,0709,121;%%
%  %\cite{Brown:2008zq}
%\bibitem{Brown:2008zq}
  C.~M.~Brown and O.~DeWolfe,
  ``Nonsupersymmetric brane vacua in stabilized compactifications,''
  JHEP {\bf 0901}, 039 (2009)
  [arXiv:0806.4399 [hep-th]].
%  %%CITATION = JHEPA,0901,039;%%

 \bibitem{FUL1} T. Fulton and F. Rohrlich,
 "Classical radiation from a uniformly accelerated charge," Ann. Phys. (N.Y.) 9, 499 (1960); 
 F. Rohrlich, Nuovo Cimento 21, 802 (1961).
 
\bibitem{SCH} Schwinger, J. S. (1998). 
{\em Classical electrodynamics.} Reading, Mass: Perseus Books. 

 %\cite{Boulware:1979qj}
\bibitem{Boulware}
  D.~G.~Boulware,
  ``Radiation From A Uniformly Accelerated Charge,''
  Annals Phys.\  {\bf 124}, 169 (1980).
  %%CITATION = APNYA,124,169;%%

\bibitem{BondiGold}
  H.~Bondi and T.~Gold,
  ``The Field Of A Uniformly Accelerated Charge, With Special Reference To The Problem Of Gravitational Acceleration,''
  Proc.\ Roy.\ Soc.\ Lond.\  A {\bf 229}, 416 (1955).
  %%CITATION = PRSLA,A229,416;%%

%\cite{Parrott:1997cp}
\bibitem{Parrott:1997cp}
  S.~Parrott,
  ``Radiation from a charge uniformly accelerated for all time,''
  Gen.\ Rel.\ Grav.\  {\bf 29}, 1463 (1997)
  [arXiv:gr-qc/9711027].
  %%CITATION = GRGVA,29,1463;%%


%\cite{Candelas:1983gz}
\bibitem{Candelas}
  P.~Candelas and D.~w.~Sciama,
  ``Is There A Quantum Equivalence Principle?,''
  Phys.\ Rev.\  D {\bf 27}, 1715 (1983).
  %%CITATION = PHRVA,D27,1715;%%  

 %\cite{Kinnersley:1970zw}
\bibitem{Kinnersley:1970zw}
  W.~Kinnersley and M.~Walker,
  ``Uniformly accelerating charged mass in general relativity,''
  Phys.\ Rev.\  D {\bf 2} (1970) 1359.
  %%CITATION = PHRVA,D2,1359;%%
 %\cite{Podolsky:2000pp}
\bibitem{Podolsky:2000pp}
  J.~Podolsky and J.~B.~Griffiths,
  ``Uniformly accelerating black holes in a de Sitter universe,''
  Phys.\ Rev.\  D {\bf 63}, 024006 (2001)
  [arXiv:gr-qc/0010109].
  %%CITATION = PHRVA,D63,024006;%%
  
 \bibitem{NIKI} A. Nikishov and V. I. Ritus, Soviet Physics JETP, Volume 29, Number 6, 1093 (1969)
 
 
 
\bibitem{hubble} S.  Eidelman et al., Phys.Lett.B592, 1 (2004); 
 O.~Lahav and A.~R.~Liddle,
  ``The cosmological parameters 2005,''
  arXiv:astro-ph/0601168.
  %%CITATION = ASTRO-PH/0601168;%%

 
 \bibitem{allenandromano}  
 B.~Allen and J.~D.~Romano,
  ``Detecting a stochastic background of gravitational radiation: Signal
  processing strategies and sensitivities,''
  Phys.\ Rev.\  D {\bf 59}, 102001 (1999)
  [arXiv:gr-qc/9710117].
  %%CITATION = PHRVA,D59,102001;%%
 

 \bibitem{Drasco:2002yd}
  S.~Drasco and E.~E.~Flanagan,
  ``Detection methods for non-Gaussian gravitational wave stochastic backgrounds,''
  Phys.\ Rev.\  D {\bf 67}, 082003 (2003)
  [arXiv:gr-qc/0210032].
  %%CITATION = PHRVA,D67,082003;%%
 


\bibitem{cutlerflanagan}  
 C.~Cutler and E.~E.~Flanagan,
  ``Gravitational waves from merging compact binaries: How accurately can one
  extract the binary's parameters from the inspiral wave form?,''
  Phys.\ Rev.\  D {\bf 49}, 2658 (1994)
  [arXiv:gr-qc/9402014].
  %%CITATION = PHRVA,D49,2658;%%

\bibitem{ligosens} A.  Lazzarini, R.  Weiss (1996), LIGO technical report, 
LIGO-E950018-02; A.  Abramovici, et al.  Science 256 (1992) 325;
"Proposal to the National Science Foundation -- The Construction,
Operation, and Supporting Research and Development of a Laser
Interferometer Gravitational-Wave Observatory", December 1989, Thorne,
Drever, Weiss, amd Raab, PHY-9210038.
  

%\cite{Hughes:2001ya}
\bibitem{Hughes:2001ya}
  S.~A.~Hughes,
  ``Untangling the merger history of massive black holes with LISA,''
  Mon.\ Not.\ Roy.\ Astron.\ Soc.\  {\bf 331}, 805 (2002)
  [arXiv:astro-ph/0108483].
  %%CITATION = MNRAA,331,805;%%

%\cite{Barack:2004wc}
\bibitem{Barack:2004wc}
  L.~Barack and C.~Cutler,
  ``Confusion noise from LISA capture sources,''
  Phys.\ Rev.\  D {\bf 70}, 122002 (2004)
  [arXiv:gr-qc/0409010].
  %%CITATION = PHRVA,D70,122002;%%

  
  
  %\cite{Caldwell:1991jj} %BBN bound
\bibitem{Caldwell:1991jj}
  R.~R.~Caldwell and B.~Allen,
  ``Cosmological Constraints On Cosmic String Gravitational Radiation,''
  Phys.\ Rev.\  D {\bf 45}, 3447 (1992).
  %%CITATION = PHRVA,D45,3447;%%

    %\cite{Abbott:2006zx}
\bibitem{Abbott:2006zx}
  B.~Abbott {\it et al.}  [LIGO Collaboration],
  ``Searching for a stochastic background of gravitational waves with LIGO,''
  Astrophys.\ J.\  {\bf 659}, 918 (2007)
  [arXiv:astro-ph/0608606].
  %%CITATION = ASJOA,659,918;%%

  
 
%\cite{Coleman:1978ae}
\bibitem{Coleman:1978ae}
  S.~R.~Coleman,
  ``The uses of instantons,''
  Subnucl.\ Ser.\  {\bf 15}, 805 (1979).
  %%CITATION = SUSEE,15,805;%%
  
  
  %\cite{Misner:1974qy}
\bibitem{MTW}
  C.~W.~Misner, K.~S.~Thorne and J.~A.~Wheeler,
  ``Gravitation,''
%\href{http://www.slac.stanford.edu/spires/find/hep/www?irn=6627595}{SPIRES entry}
{\it  San Francisco 1973, 1279p}


 

\end{thebibliography}
\end{document}